\documentclass[12pt,a4paper]{article}
\usepackage{amsmath,caption,amsfonts,setspace,mathrsfs,mathtools}
\usepackage[top=1.1in, bottom=1.1in, left=1.1in, right=1.1in]{geometry}
\usepackage[round]{natbib}
\usepackage{color,soul}

\DeclareCaptionStyle{italic}[justification=centering]{labelfont={bf},textfont={it},labelsep=colon}
\usepackage{graphicx,psfrag,epsf,textcomp}
\usepackage{enumerate, dsfont}
\usepackage{natbib}
\usepackage{url,xcolor} 
\usepackage{booktabs, subfig, bm, paralist,mathpazo,tikz,todonotes,longtable,microtype}

\usepackage[pdftex,colorlinks=true]{hyperref}
\definecolor{darkblue}{rgb}{0,0,.6}
\hypersetup{citecolor=darkblue,linkcolor=darkblue,urlcolor=darkblue}
\newcommand{\argmax}{\operatornamewithlimits{argmax}}

\newcommand{\blind}{0}
\newcommand{\X}{\mathcal{X}}
\definecolor{a0}{rgb}{0.0, 0.5, 0.0}
\definecolor{bistre}{rgb}{0.24, 0.17, 0.12}
\definecolor{amethyst}{rgb}{0.6, 0.4, 0.8}
\definecolor{blue-violet}{rgb}{0.54, 0.17, 0.89}
\definecolor{Rcolor}{RGB}{150,160,190}
\definecolor{blush}{rgb}{0.87, 0.36, 0.51}
\definecolor{brightturquoise}{rgb}{0.03, 0.91, 0.87}
\definecolor{burntorange}{rgb}{0.8, 0.33, 0.0}

\graphicspath{{plots/}}
\DeclareMathOperator*{\argmin}{\arg\!\min}
\newsavebox\CBox
\def\textBF#1{\sbox\CBox{#1}\resizebox{\wd\CBox}{\ht\CBox}{\textbf{#1}}}

\begin{document}

\def\spacingset#1{\renewcommand{\baselinestretch}{#1}\small\normalsize} \spacingset{1}

\if0\blind
{
  \title{\bf Dynamic principal component regression for forecasting functional time series in a group structure}
  \author{Han Lin Shang\footnote{Postal address: Research School of Finance, Actuarial Studies and Statistics, Level 4, Building 26C, Kingsley Street, Australian National University, ACT 2601, Australia; Email: hanlin.shang@anu.edu.au; Telephone: +61(2) 612 50535; Fax: +61(2) 612 50087; ORCID: \url{https://orcid.org/0000-0003-1769-6430}} \\
	Australian National University}
  \maketitle
} \fi

\if1\blind
{
  \title{\bf Dynamic principal component regression for forecasting functional time series in a group structure}
  \maketitle
} \fi

\bigskip

\begin{abstract}

When generating social policies and pricing annuity at national and subnational levels, it is essential both to forecast mortality accurately and ensure that forecasts at the subnational level add up to the forecasts at the national level. This has motivated recent developments in forecasting functional time series in a group structure, where static principal component analysis is used. In the presence of moderate to strong temporal dependence, static principal component analysis designed for independent and identically distributed functional data may be inadequate. Thus, through using the dynamic functional principal component analysis, we consider a functional time series forecasting method with static and dynamic principal component regression to forecast each series in a group structure. Through using the regional age-specific mortality rates in Japan obtained from the \cite{JMD18}, we investigate the point and interval forecast accuracies of our proposed extension, and subsequently make recommendations.

\vspace{.1in}
\end{abstract}

\def\spacingset#1{\renewcommand{\baselinestretch}{#1}\small\normalsize} \spacingset{1}
\spacingset{1.33}

\newpage

\section{Introduction}

In recent years, rapidly aging populations and increased longevity in many developed countries have been a growing concern for governments and societies. These concerns are centered on the sustainability of pensions and health and aged-care systems, especially given the increased longevity among populations. The importance of mortality has resulted in a surge of interest among government policymakers and urban and regional planners in accurately modeling and forecasting age-specific mortality at national and regional levels. An improvement in the forecast accuracy of mortality would be greatly beneficial for policy decisions regarding the allocation of current and future resources to regions. Further, future mortality rates represent input to determine the life, fixed-term, and delayed annuity prices, and thus, they are of great interest to the insurance and pension industries.

Many statistical methods have been proposed to forecast age-specific mortality rates and their associated life expectancies \citep[c.f.][for earlier reviews]{Booth06, BT08, CDE04, GK08, TB14}. Of these, a significant milestone in demographic forecasting was the work of \cite{LC92}. The strengths of the Lee--Carter (LC) method are simplicity and robustness in cases where age-specific log mortality rates have linear trends \citep{BHT+06}.  The weakness of the LC method is that it attempts to capture mortality patterns using one principal component and its associated scores. To overcome this deficiency, the LC method has been extended and modified in two streams. From the perspective of the time series of a matrix, \cite{BMS02}, \cite{RH03}, and \cite{CBD06, CBD+09} proposed the use of more than one component in the LC method to model age-specific mortality. \cite{CBD06} used a logistic transformation to model the relationship between the probability of death and age observed over time, and then \cite{CBD+09} extended this model by incorporating the cohort effect. \cite{GK08} and \cite{WSB+15} considered the Bayesian paradigm for parameter estimation and forecasting in the LC method. 

A common drawback of these works is that a static (functional) principal component analysis is often used to decompose a time series of data matrices or curves. For the case of independent data, the eigendecomposition is usually conducted on (sample) variance function, seeking an optimal linear combination of observations that could maximize the variance function \citep[c.f.,][]{RS05}. When the temporal dependence in a functional time series is moderate or strong, a functional version of auto-correlation function at various lags may be significantly above the constant threshold. In this case, the extracted principal components from an estimated variance operator are not consistent because of temporal dependence, leading to erroneous estimators and loss of sample dynamic information. To rectify this issue, it is more sensible to use a long-run covariance function, which accounts for auto-correlation \citep[c.f.,][]{HK12, HKH15}. Long-run covariance includes the variance as a component, yet also measures the auto-covariance at various positive and negative lags. From an estimated long-run covariance, we apply a dynamic approach to extract principal components. 

This paper makes three contributions. First, we demonstrate the improvement of point and interval forecast accuracies that the dynamic functional principal component analysis (DFPCA) entails when compared with a functional time series method based on the static functional principal component analysis (FPCA) for modeling and forecasting individual age-specific mortality series at a short- to moderate-term forecast horizon. Second, in a group structure, we demonstrate the improvement of point and interval forecast accuracies of the DFPCA produces when compared with the functional time series method based on the statistic FPCA. Third, we present a procedure for constructing pointwise prediction intervals. 

The remainder of this paper is given as follows. In Section~\ref{sec:2}, we present age-specific mortality rates obtained from the \cite{JMD18}. Section~\ref{sec:3} describes a kernel sandwich estimator for estimating long-run covariance. Based on the estimated long-run covariance, we perform an eigendecomposition that extracts dynamic principal components and their associated scores. In Section~\ref{sec:4}, we adopt the procedure of \cite{ANH15} to construct prediction intervals. In Section~\ref{sec:5}, we revisit two methods considered by \cite{SH17b} to reconcile forecasts in a group structure. In the group structure, we also find that the dynamic functional principal component decomposition produces more accurate forecasts in comparison with the static functional principal component decomposition. In Sections~\ref{sec:6} and~\ref{sec:int}, we evaluate and compare the short- to the moderate-term point and interval forecast accuracies between the FPCA and DFPCA within a group structure. The conclusions are presented in Section~\ref{sec:7}.

\section{Japanese age-specific mortality rates for 47 prefectures}\label{sec:2}

We study Japanese age-specific mortality rates from 1975 to 2016, obtained from the \cite{JMD18}. We consider ages from zero to 99 in single years of age, while the last age group contains all ages at and above 100 years. In Table~\ref{tab:1}, we present a data group structure where each row denotes a level of disaggregation. The top level displays the total age-specific mortality rates for Japan. The overall mortality rates can be split into different sex, region, or prefecture groups. There are eight regions in Japan, which contain a total of 47 prefectures. The most disaggregated data arise when we consider the mortality rates for each prefecture and each sex, giving a total of $47\times 2 = 94$ series. In total, across all levels of disaggregation, there are 168 series.

\begin{table}[!htbp]
\centering
\tabcolsep 0.58in
\caption{Group structure of Japanese mortality rates.}\label{tab:1}
\begin{tabular}{@{}lr@{}}
\toprule
Level & Number of series \\
\midrule
Japan & 1 \\
Sex & 2 \\
Region & 8 \\
Sex $\times$ Region & 16 \\
Prefecture & 47 \\
Sex $\times$ Prefecture & 94 \\
\midrule
Total & 168 \\
\bottomrule
\end{tabular}
\end{table}

\begin{figure}[!htbp]
\centering
\subfloat[Observed female mortality in Japan]
{\includegraphics[width=8.5cm]{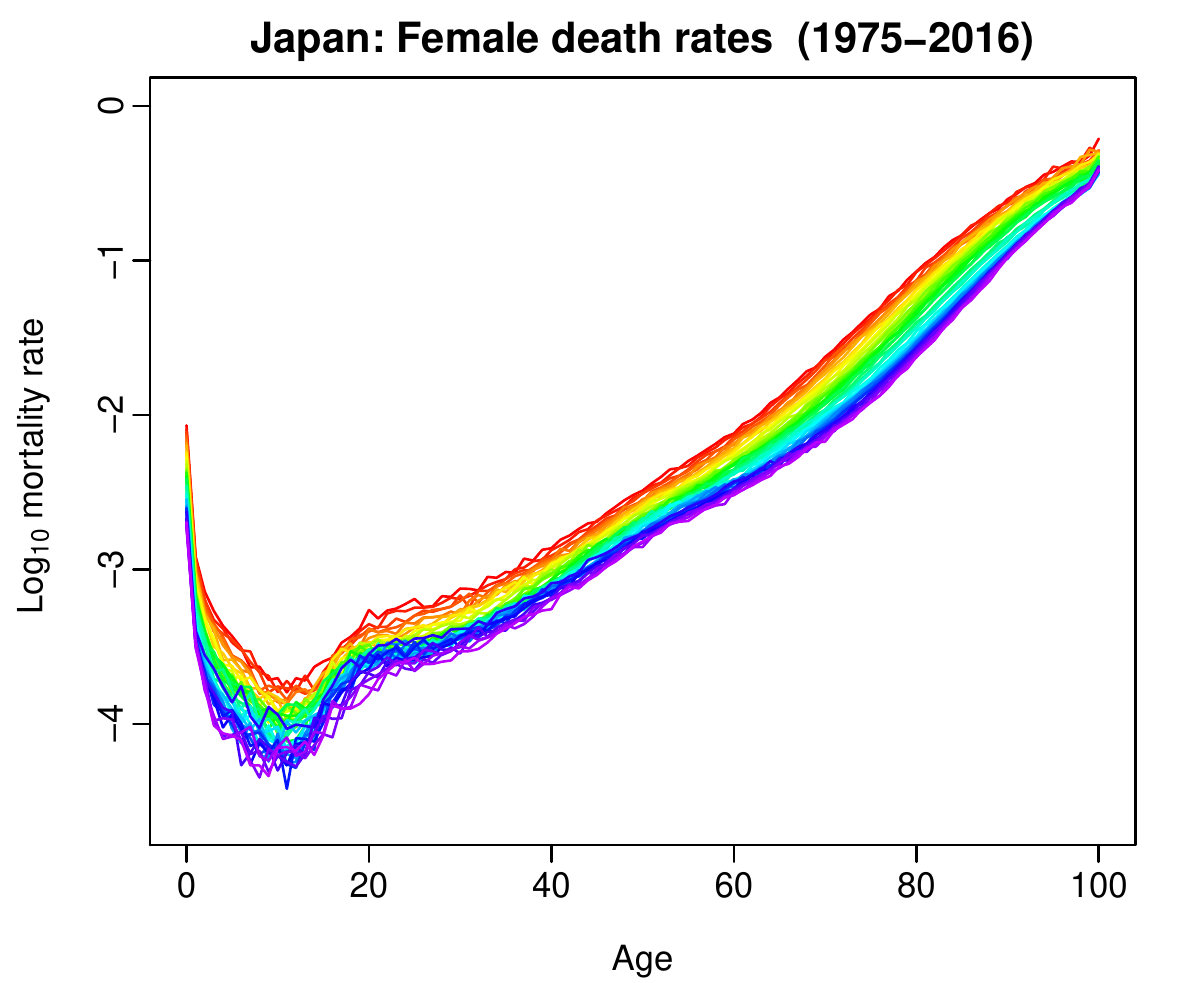}\label{fig:1a}}
\qquad
\subfloat[Smoothed female mortality in Japan]
{\includegraphics[width=8.5cm]{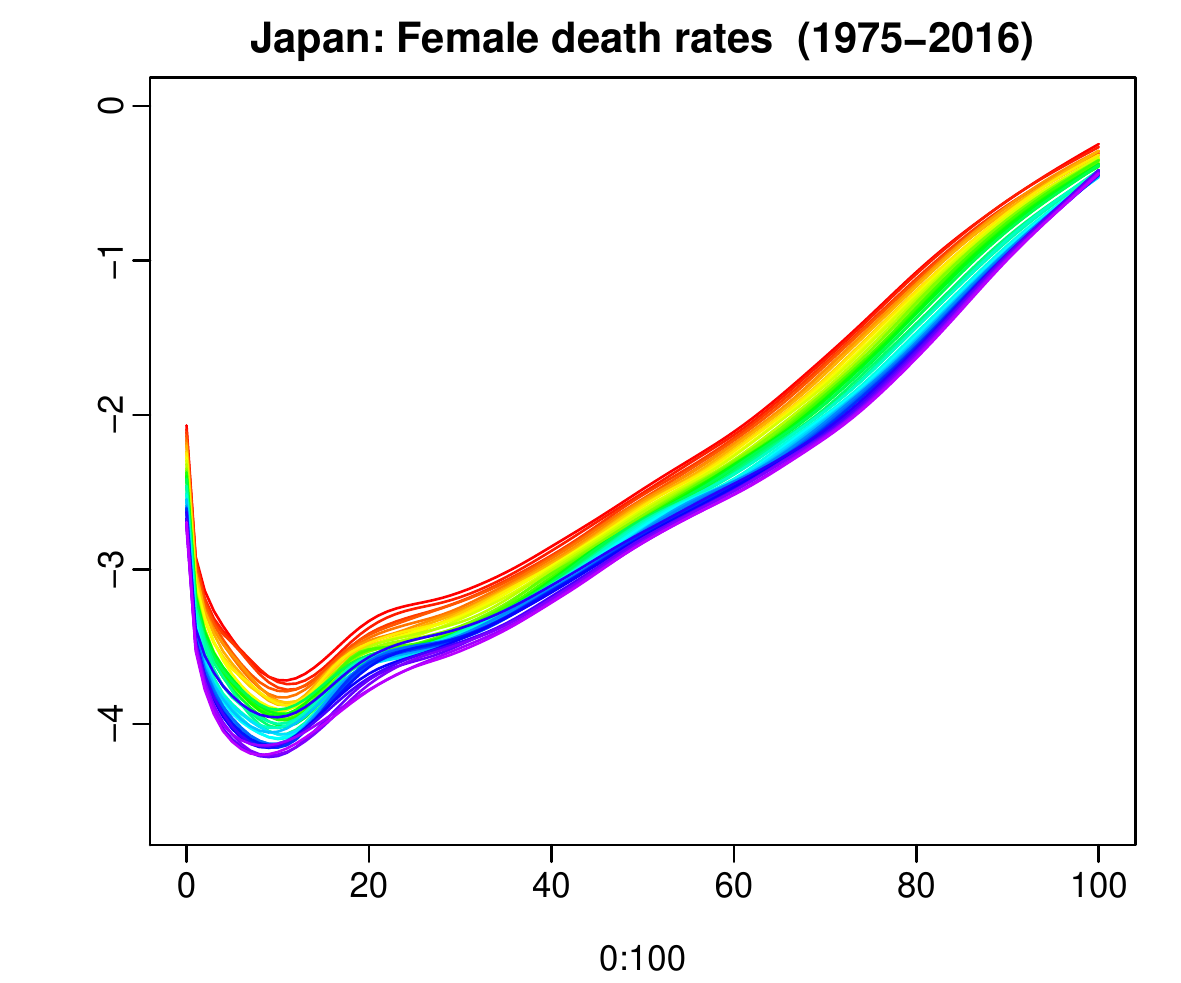}\label{fig:1b}}
\\
\subfloat[Observed female mortality in Hokkaido]
{\includegraphics[width=8.5cm]{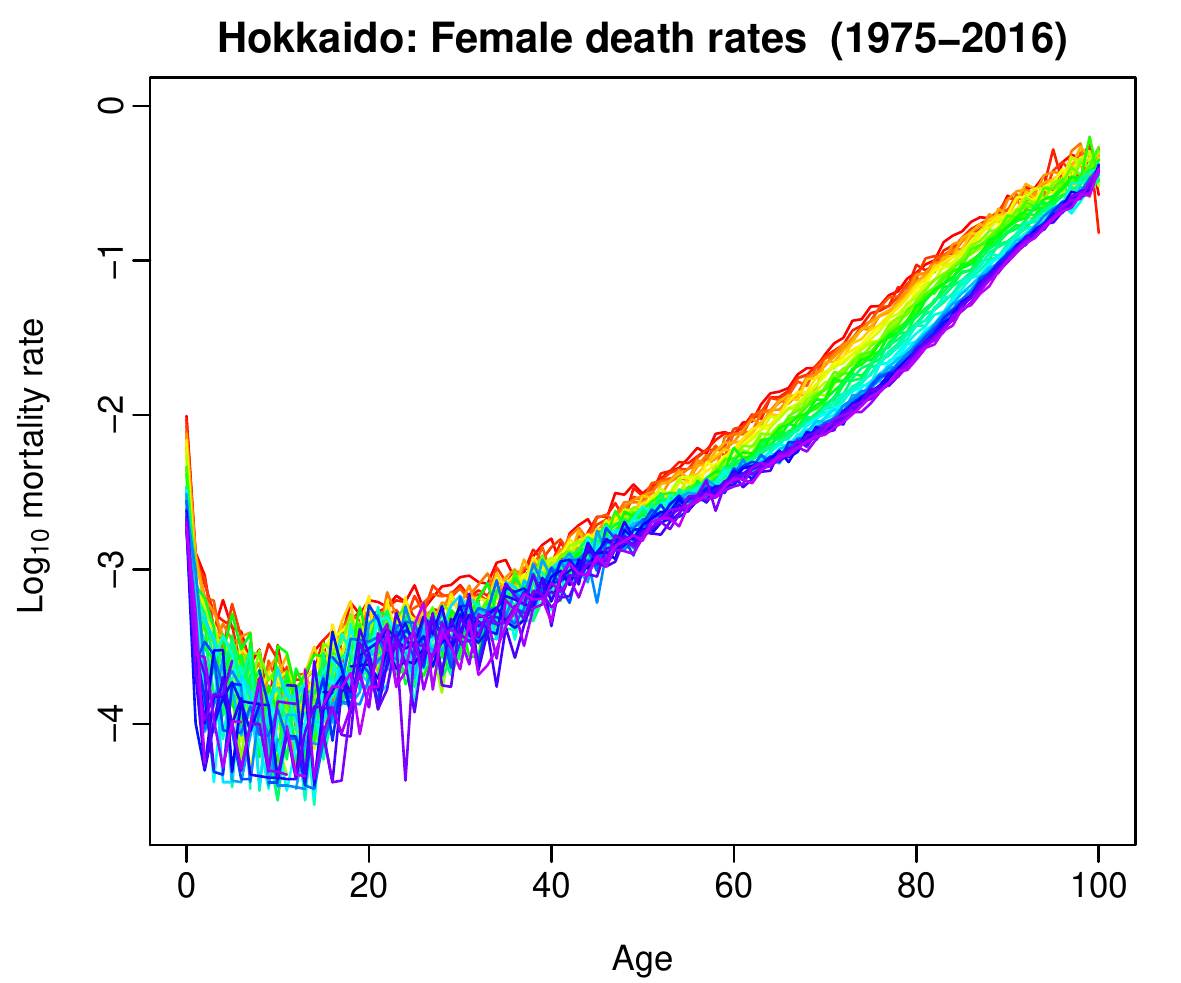}\label{fig:1c}}
\qquad
\subfloat[Smoothed female mortality in Hokkaido]
{\includegraphics[width=8.5cm]{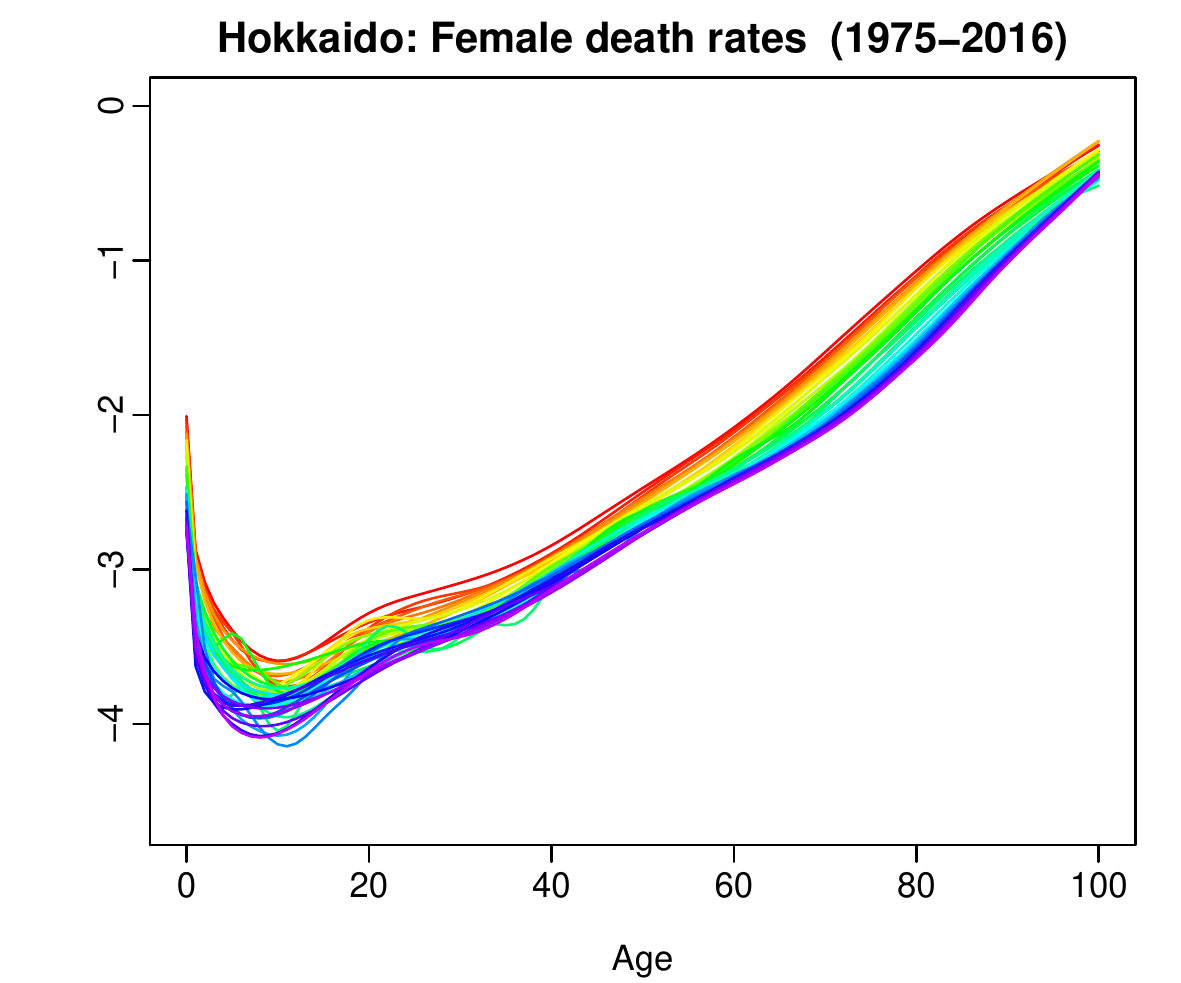}\label{fig:1d}}
\caption{Functional time series graphical displays. Logarithm base 10 has a simpler interpretation than the natural logarithm.}\label{fig:1}
\end{figure}

To view the time-series evolution of the Japanese age-specific mortality, we present functional time series plots for the logarithm base 10 of female mortality rates in Japan and Hokkaido in Figures~\ref{fig:1a} and~\ref{fig:1c}, while the functional time series plots for the smoothed data are shown in Figures~\ref{fig:1b} and~\ref{fig:1d}. To smooth these functional time series, we use a penalized regression spline with monotonic constraint, where the monotonicity is imposed for ages at and above 65 \citep[see, e.g.,][]{HU07}.

As shown in Figure~\ref{fig:1}, by analyzing the changes in mortality as a function of both age $x$ and year $t$, it is evident that mortality rates have displayed a gradual decline over the years. Mortality rates dip from their early childhood high, climb in the teen years, stabilize in the early twenties, and then steadily increase with age. For both females and males, log mortality rates have been declining over the years, especially in the younger and older ages. Further, the observed Japanese national female mortality rates contain less noise than the observed Hokkaido female mortality rates. Thus, the nonparametric smoothing technique is more useful for the Hokkaido series (noisier) than for the Japanese national series (less noisy).

\section{Dynamic functional principal component analysis}\label{sec:3}

The estimation of long-run covariance has not been considered in actuarial science, except for the work of \cite{Shang18}. Using mortality rates in developed countries, \cite{Shang18} found that the dynamic principal components extracted from the long-run covariance function better capture temporal dependency than do the static principal components extracted from the variance function.

Using the bandwidth estimation procedure of \cite{RS17} and the kernel sandwich estimator, we obtain an estimate of long-run covariance $\widehat{C}_{h, q}(u, s)$. We then apply functional principal component decomposition to the estimated long-run covariance and obtain the functional principal components and their associated scores. Through employing Karhunen--Lo\`{e}ve expansion, a stochastic process, $\X$, can be expressed as
\begin{equation*}
\X(u) = \mu(u) + \sum^{\infty}_{j=1}\beta_j\phi_j(u),
\end{equation*}
where $\mu(u)$ denotes the mean function, $\X^c(u) = \X(u) - \mu(u)$, and $\beta_j$ is an uncorrelated random variable with zero mean and unit variance. The principal component scores, $\beta_j$, are given by the projection of $\X(u) - \mu(u)$ in the direction of the $j$\textsuperscript{th} eigenfunction, $\phi_j$--that is, $\beta_j =\langle \X(u)-\mu(u), \phi_j(u)\rangle$.

The DFPCA summarizes the primary features of an infinite-dimensional object by its finite vital elements, and forms a base of dynamic functional principal component regression in Section~\ref{sec:DFPCR}. For theoretical, methodological, and applied aspects of DFPCA, consult the articles by \cite{HKH15} and \cite{RS17}.

\subsection{Dynamic functional principal component regression}\label{sec:DFPCR}

Conditional on the observed time series of smooth functions, $\bm{\X}(u)$, the $h$-step-ahead forecast of $\X_{n+h}(u)$ can be given by
\begin{align*}
\widehat{\X}_{n+h|n}(u) = \text{E}\left[\X_{n+h}(u)|\bm{\X}(u), \bm{\Phi}(u)\right] = \overline{\X}(u) + \sum^J_{j=1}\widehat{\beta}_{n+h|n,j}\widehat{\phi}_j(u),
\end{align*}
where $\overline{\X}(u)$ denotes a mean function; $\bm{\Phi}(u) = \{\widehat{\phi}_1(u),\dots,\widehat{\phi}_J(u)\}$ denotes a set of functional principal components; $\widehat{\beta}_{n+h|n,j}$ denotes the forecast principal component score obtained from a univariate time series forecasting method for $j$\textsuperscript{th} component; and $J$ denotes the retained number of components.

The number of components, $J$, can be selected as the minimum that reaches a certain level of the cumulative proportion of variance (CPV) explained by the first $J$ leading components. Thus:
\begin{equation}
J_{\text{CPV}} = \argmin_{J: 1\leq J\leq n}\left\{\sum^J_{j=1}\widehat{\lambda}_j\Big/\sum^n_{j=1}\widehat{\lambda}_j\geq \delta\right\}, \label{eq:2}
\end{equation}
where $\widehat{\lambda}_j$ denotes the estimated $j$\textsuperscript{th} eigenvalue, and $\delta = 85\%$ \citep[see, e.g.,][p.41]{HK12}.

The criterion in~\eqref{eq:2} remains fixed with sample size $n$, although the optimal order of $J$ should increase to infinity with $n$. One way to accommodate such an aspect is to use the criterion of \cite{HK15}, which explicitly incorporates the behavior of the estimated eigenvalues $(\widehat{\lambda}_1, \widehat{\lambda}_2, \dots, )$. The number of components, $J$, is selected by the rule
\begin{equation*}
J_{\text{HK}} = \argmax_{J: 1\leq J \leq n}\left\{\frac{\widehat{\lambda}_1}{\widehat{\lambda}_J}\leq \frac{\sqrt{n}}{\log_{10}(n)}\right\}.
\end{equation*}
As pointed out by \cite{HBY13}, there is little influence on forecast accuracy if we select the number of components greater than the optimal (yet unknown) one. Thus, the selected value of $J$ is the maximum between the ones selected by the CPV criterion and the criterion of \cite{HK15}--that is:
\begin{equation}
J = \max\{J_{\text{CPV}}, J_{\text{HK}}\}. \label{eq:3}
\end{equation}

\subsection{Constructing pointwise prediction intervals}\label{sec:4}

To construct pointwise prediction intervals, we adopt the method of \cite{ANH15}. The method can be summarized in the following steps:
\begin{enumerate}
\item Using all observations, we compute the $J$-variate score vectors ($\bm{\widehat{\beta}}_1,\dots,\bm{\widehat{\beta}}_J$) and the sample (dynamic) functional principal components $[\widehat{\phi}_1(u),\dots,\widehat{\phi}_J(u)]$. We then compute in-sample point forecasts
\begin{equation*}
\widehat{\X}_{\zeta+h|\zeta}(u)=\widehat{\beta}_{\zeta+h|\zeta,1}\widehat{\phi}_1(u) + \cdots + \widehat{\beta}_{\zeta+h|\zeta,J}\widehat{\phi}_J(u),
\end{equation*}
where $(\widehat{\beta}_{\zeta+h|\zeta,1},\dots,\widehat{\beta}_{\zeta+h|\zeta,J})$ are the elements of the $h$-step-ahead prediction obtained from $(\widehat{\bm{\beta}}_1,\dots,\widehat{\bm{\beta}}_J)$ by a univariate time series forecasting method for $\zeta\in \{J,\dots,n-h\}$.
\item With the in-sample point forecasts, we compute the in-sample point forecast errors
\begin{equation}
\widehat{\epsilon}_{\omega}(u) = \X_{\zeta+h}(u)-\widehat{\X}_{\zeta+h|\zeta}(u), \label{eq:4}
\end{equation}
where $\omega=1,2,\dots, M$ and $M=n-h-J+1$.
\item Based on the in-sample forecast errors, we take quantiles to obtain lower and upper prediction intervals, denoted by $\gamma^{\text{lb}}(u)$ and $\gamma^{\text{ub}}(u)$, respectively, for a nominal coverage probability. Then, we seek a tuning parameter, $\psi_{\alpha}$, such that $\alpha\times 100\%$ of the residual functions satisfies:
\begin{equation*}
\psi_{\alpha}\times \gamma^{\text{lb}}(u) \leq \widehat{\epsilon}_{\omega}(u)\leq \psi_{\alpha}\times \gamma^{\text{ub}}(u).
\end{equation*}
Based on the law of large numbers, the residuals $\left[\widehat{\epsilon}_1(u),\dots,\widehat{\epsilon}_{M}(u)\right]$ are expected to be approximately stationary and satisfy:
\begin{small}
\begin{equation*}
\frac{1}{M}\sum^M_{\omega=1}\mathds{1}[\psi_{\alpha}\times \gamma^{\text{lb}}(u)\leq \widehat{\epsilon}_{\omega}(u)\leq \psi_{\alpha}\times \gamma^{\text{ub}}(u)] \approx \text{Pr}[\psi_{\alpha}\times \gamma^{\text{lb}}(u)\leq \X_{n+h}(u) - \widehat{\X}_{n+h|n}(u)\leq \psi_{\alpha}\times \gamma^{\text{ub}}(u)],
\end{equation*}
\end{small}
where $\psi_{\alpha}$ calibrates the difference between empirical and nominal coverage probabilities.
\end{enumerate}
Step 3 can be extended to pointwise prediction intervals, where we have $\psi_{\alpha}=1$, such that $\alpha\times 100\%$ of the residual data points satisfy:
\begin{equation}
\psi_{\alpha}\times \gamma^{\text{lb}}(u_i)\leq \widehat{\epsilon}_{\omega}(u_i)\leq \psi_{\alpha}\times \gamma^{\text{ub}}(u_i).\label{eq:5}
\end{equation}
Then, the $h$-step-ahead pointwise prediction intervals of $\X_{n+h}(u_i)$ are given as:
\begin{equation*}
\left(\widehat{\X}_{n+h|n}(u_i)+\psi_{\alpha}\times \gamma^{\text{lb}}(u_i), \widehat{\X}_{n+h|n}(u_i)+\psi_{\alpha}\times \gamma^{\text{ub}}(u_i)\right),
\end{equation*}
where $i=1,\dots,101$ denotes the discretized data points.

\section{Grouped functional time series forecasting}\label{sec:5}

\subsection{Notation}

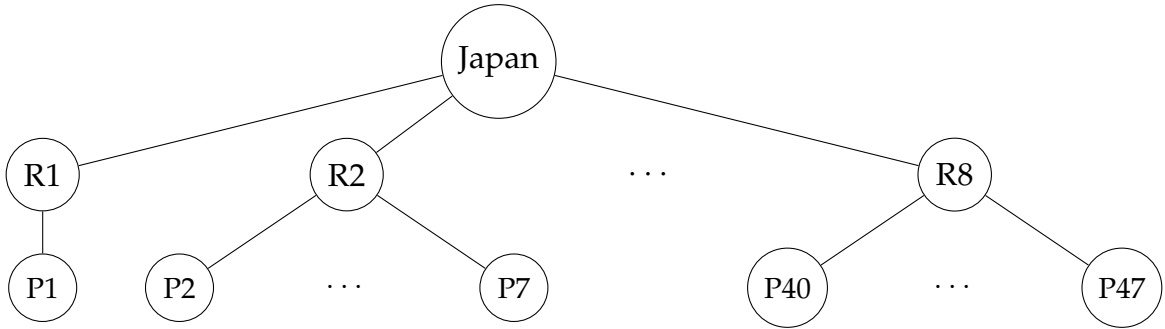
\begin{figure}[!htb]
\centering\begin{tikzpicture}
\tikzstyle{every node}=[minimum size = 8mm]
\tikzstyle[level distance=10cm] \tikzstyle[sibling distance=40cm]
\tikzstyle{level 3}=[sibling distance=16mm,font=\footnotesize]
\tikzstyle{level 2}=[sibling distance=22mm,font=\small]
\tikzstyle{level 1}=[sibling distance=40mm,font=\normalsize]
\node[circle,draw]{Japan}
   child {node[circle,draw] {R1}
   	     child {node[circle,draw] {P1}}}
   child {node[circle,draw] {R2}
   		child {node[circle,draw] {P2}}
      child {node {$\cdots$}edge from parent[draw=none]}
		child {node[circle,draw] {P7}}
		}
  child {node {$\cdots$}edge from parent[draw=none]}
   child {node[circle,draw] {R8}
   		child{node[circle,draw] {P40}}
	     child{node {$\cdots$}edge from parent[draw=none]}
  	child{node[circle,draw] {P47}}
 };
\end{tikzpicture}
\medskip
\caption{Japanese geographical tree diagram, with eight regions and 47 prefectures--each node has female, male, and total age-specific mortality rates.}\label{fig:2}
\end{figure}

For ease of explanation, we introduce the notation using the Japanese example presented in Section~\ref{sec:2}. The Japanese data follow a multi-level geographical group structure coupled with a sex grouping variable. The group structure is shown in Figure~\ref{fig:2}. Japan is split into eight regions, which in turn can be divided into 47 prefectures.
\medskip

The data can also be split by sex. Thus, each of the nodes in the geographical hierarchy can also be split into both males and females. We refer to a particular disaggregated series using the notation $\text{X}\ast \text{S}$, meaning the geographical area $\text{X}$ and the sex $\text{S}$, where $\text{X}$ can take the values shown in Figure~\ref{fig:2} and $\text{S}$ can take values M (males), F (females), or T (total). For example, $\text{R1}\ast \text{F}$ denotes females in Region 1, $\text{P1}\ast \text{T}$ denotes females and males in Prefecture 1, $\text{Japan}\ast \text{M}$ denotes males in Japan, and so on.

Let $\text{E}_{\text{X}\ast \text{S}, t}(u)$ denote the exposure-at-risk for series $\text{X}\ast\text{S}$ in year $t$ and age $u$, and let $\text{D}_{\text{X}\ast \text{S}, t}(z)$ be the number of deaths for series $\text{X}\ast \text{S}$ in year $t$ and age $u$. The age-specific mortality rate is expressed as
\begin{equation*}
\text{R}_{\text{X}\ast\text{S}, t}(u) = \text{D}_{\text{X}\ast\text{S}, t}(u)/\text{E}_{\text{X}\ast\text{S}, t}(u).
\end{equation*}
To simplify expressions, we drop the age argument of $u$. Then, for a given age, we can write
\[
\hspace{-.2in} \underbrace{ \left[
\begin{footnotesize}
\begin{array}{l}
\text{R}_{\text{Japan*T},t} \\
\text{R}_{\textcolor{red}{\text{Japan*F},t}} \\
\text{R}_{\textcolor{red}{\text{Japan*M},t}} \\
\text{R}_{\textcolor{a0}{\text{R1*T},t}} \\
\text{R}_{\textcolor{a0}{\text{R2*T},t}} \\
\vdots \\
\text{R}_{\textcolor{a0}{\text{R8*T},t}} \\
\text{R}_{\textcolor{blue-violet}{\text{R1*F},t}} \\
\text{R}_{\textcolor{blue-violet}{\text{R2*F},t}} \\
\vdots \\
\text{R}_{\textcolor{blue-violet}{\text{R8*F},t}} \\
\text{R}_{\textcolor{burntorange}{\text{R1*M},t}} \\
\text{R}_{\textcolor{burntorange}{\text{R2*M},t}} \\
\vdots \\
\text{R}_{\textcolor{burntorange}{\text{R8*M},t}} \\
\text{R}_{\textcolor{blue}{\text{P1*T},t}} \\
\text{R}_{\textcolor{blue}{\text{P2*T},t}} \\
\vdots \\
\text{R}_{\textcolor{blue}{\text{P47*T},t}} \\
\text{R}_{\textcolor{purple}{\text{P1*F},t}} \\
\text{R}_{\textcolor{purple}{\text{P1*M},t}} \\
\text{R}_{\textcolor{purple}{\text{P2*F},t}} \\
\text{R}_{\textcolor{purple}{\text{P2*M},t}} \\
\vdots \\
\text{R}_{\textcolor{purple}{\text{P47*F},t}} \\
\text{R}_{\textcolor{purple}{\text{P47*M},t}} \\ \end{array}
\end{footnotesize} \right]}_{\bm{R}_t} =
\underbrace{\left[
\begin{footnotesize}
\begin{array}{ccccccccccc}
\frac{\text{E}_{\text{P1*F},t}}{\text{E}_{\text{Japan*T},t}} & \frac{\text{E}_{\text{P1*M},t}}{\text{E}_{\text{Japan*T},t}} & \frac{\text{E}_{\text{P2*F},t}}{\text{E}_{\text{Japan*T},t}} & \frac{\text{E}_{\text{P2*M},t}}{\text{E}_{\text{Japan*T},t}}  & \frac{\text{E}_{\text{P3*F},t}}{\text{E}_{\text{Japan*T},t}} & \frac{\text{E}_{\text{P3*M},t}}{\text{E}_{\text{Japan*T},t}} & \cdots & \frac{\text{E}_{\text{P47*F},t}}{\text{E}_{\text{Japan*T},t}} & \frac{\text{E}_{\text{P47*M},t}}{\text{E}_{\text{Japan*T},t}} \\
\textcolor{red}{\frac{\text{E}_{\text{P1*F},t}}{\text{E}_{\text{Japan*F},t}}} & \textcolor{red}{0} & \textcolor{red}{\frac{\text{E}_{\text{P2*F},t}}{\text{E}_{\text{Japan*F},t}}} & \textcolor{red}{0} & \textcolor{red}{\frac{\text{E}_{\text{P3*F},t}}{\text{E}_{\text{Japan*F},t}}} & \textcolor{red}{0} & \cdots & \textcolor{red}{\frac{\text{E}_{\text{P47*F},t}}{\text{E}_{\text{Japan*F},t}}} & \textcolor{red}{0} \\
\textcolor{red}{0} & \textcolor{red}{\frac{\text{E}_{\text{P1*M},t}}{\text{E}_{\text{Japan*M},t}}}  & \textcolor{red}{0} & \textcolor{red}{\frac{\text{E}_{\text{P2*M},t}}{\text{E}_{\text{Japan*M},t}}} & \textcolor{red}{0} & \textcolor{red}{\frac{\text{E}_{\text{P3*M},t}}{\text{E}_{\text{Japan*M},t}}} & \cdots & \textcolor{red}{0} & \textcolor{red}{\frac{\text{E}_{\text{P47*M},t}}{\text{E}_{\text{Japan*M},t}}} \\
\textcolor{a0}{\frac{\text{E}_{\text{P1*F},t}}{\text{E}_{\text{R1,T},t}}} & \textcolor{a0}{\frac{\text{E}_{\text{P1*M},t}}{\text{E}_{\text{R1,T},t}}} & \textcolor{a0}{0} & \textcolor{a0}{0} & \textcolor{a0}{0} & \textcolor{a0}{0} & \cdots  & \textcolor{a0}{0} & \textcolor{a0}{0} \\
\textcolor{a0}{0} & \textcolor{a0}{0} & \textcolor{a0}{\frac{\text{E}_{\text{P2*F},t}}{\text{E}_{\text{R2,T},t}}} & \textcolor{a0}{\frac{\text{E}_{\text{P2*M},t}}{\text{E}_{\text{R2,T},t}}} & \textcolor{a0}{\frac{\text{E}_{\text{P3*F},t}}{\text{E}_{\text{R2,T},t}}} & \textcolor{a0}{\frac{\text{E}_{\text{P3*M},t}}{\text{E}_{\text{R2,T},t}}} & \cdots & \textcolor{a0}{0} & \textcolor{a0}{0} \\
\vdots & \vdots & \vdots & \vdots & \vdots & \vdots & \cdots & \vdots & \vdots \\
\textcolor{a0}{0} & \textcolor{a0}{0} & \textcolor{a0}{0} & \textcolor{a0}{0} & \textcolor{a0}{0} & \textcolor{a0}{0} & \cdots & \textcolor{a0}{\frac{\text{E}_{\text{P47*F},t}}{\text{E}_{\text{R8,T},t}}} & \textcolor{a0}{\frac{\text{E}_{\text{P47*M},t}}{\text{E}_{\text{R8,T},t}}} \\
\textcolor{blue-violet}{\frac{\text{E}_{\text{P1*F},t}}{\text{E}_{\text{R1,F},t}}} & \textcolor{blue-violet}{0} & \textcolor{blue-violet}{0} & \textcolor{blue-violet}{0} & \textcolor{blue-violet}{0} & \textcolor{blue-violet}{0} &  \cdots & \textcolor{blue-violet}{0} & \textcolor{blue-violet}{0} \\
\textcolor{blue-violet}{0} & \textcolor{blue-violet}{0} & \textcolor{blue-violet}{\frac{\text{E}_{\text{P2*F},t}}{\text{E}_{\text{R2,F},t}}} & \textcolor{blue-violet}{0} & \textcolor{blue-violet}{\frac{\text{E}_{\text{P3*F},t}}{\text{E}_{\text{R2,F},t}}} & \textcolor{blue-violet}{0} & \cdots & \textcolor{blue-violet}{0} & \textcolor{blue-violet}{0}  \\
\vdots & \vdots & \vdots & \vdots & \vdots & \vdots & \cdots & \vdots & \vdots \\
\textcolor{blue-violet}{0} & \textcolor{blue-violet}{0}  & \textcolor{blue-violet}{0}  & \textcolor{blue-violet}{0}  & \textcolor{blue-violet}{0}  & \textcolor{blue-violet}{0}  & \cdots & \textcolor{blue-violet}{\frac{\text{E}_{\text{P47*F},t}}{\text{E}_{\text{R8,F},t}}} & \textcolor{blue-violet}{0}\\
\textcolor{burntorange}{0} & \textcolor{burntorange}{\frac{\text{E}_{\text{P1*M},t}}{\text{E}_{\text{R1,M},t}}} & \textcolor{burntorange}{0} &\textcolor{burntorange}{0} & \textcolor{burntorange}{0} & \textcolor{burntorange}{0} & \cdots & \textcolor{burntorange}{0} & \textcolor{burntorange}{0} \\
\textcolor{burntorange}{0} & \textcolor{burntorange}{0} & \textcolor{burntorange}{0} & \textcolor{burntorange}{\frac{\text{E}_{\text{P2*M},t}}{\text{E}_{\text{R2,M},t}}} & \textcolor{burntorange}{0} & \textcolor{burntorange}{\frac{\text{E}_{\text{P3*M},t}}{\text{E}_{\text{R2,M},t}}} & \cdots & \textcolor{burntorange}{0} & \textcolor{burntorange}{0} \\
\vdots & \vdots & \vdots & \vdots & \vdots & \vdots & \cdots & \vdots & \vdots \\
\textcolor{burntorange}{0} & \textcolor{burntorange}{0} & \textcolor{burntorange}{0} & \textcolor{burntorange}{0} & \textcolor{burntorange}{0} & \textcolor{burntorange}{0} & \cdots & \textcolor{burntorange}{0} & \textcolor{burntorange}{\frac{\text{E}_{\text{P47*M},t}}{\text{E}_{\text{R8,M},t}}} \\
\textcolor{blue}{\frac{\text{E}_{\text{P1*F},t}}{\text{E}_{\text{P1,T},t}}} & \textcolor{blue}{\frac{\text{E}_{\text{P1*M},t}}{\text{E}_{\text{P1,T},t}}} & \textcolor{blue}{0} & \textcolor{blue}{0} & \textcolor{blue}{0} & \textcolor{blue}{0} & \cdots & \textcolor{blue}{0} & \textcolor{blue}{0} \\
\textcolor{blue}{0} & \textcolor{blue}{0}  &  \textcolor{blue}{\frac{\text{E}_{\text{P2*F},t}}{\text{E}_{\text{P2,T},t}}} & \textcolor{blue}{\frac{\text{E}_{\text{P2*M},t}}{\text{E}_{\text{P2,T},t}}} & \textcolor{blue}{0} & \textcolor{blue}{0}  & \cdots & \textcolor{blue}{0} & \textcolor{blue}{0} \\
\vdots & \vdots & \vdots & \vdots & \vdots & \vdots & \cdots & \vdots & \vdots \\
\textcolor{blue}{0} & \textcolor{blue}{0} & \textcolor{blue}{0} & \textcolor{blue}{0} & \textcolor{blue}{0} & \textcolor{blue}{0} & \cdots & \textcolor{blue}{\frac{\text{E}_{\text{P47*F},t}}{\text{E}_{\text{P47,T},t}}} & \textcolor{blue}{\frac{\text{E}_{\text{P47*M},t}}{\text{E}_{\text{P47,T},t}}} \\
\textcolor{purple}{1} & \textcolor{purple}{0} & \textcolor{purple}{0} & \textcolor{purple}{0} & \textcolor{purple}{0} & \textcolor{purple}{0} & \cdots & \textcolor{purple}{0} & \textcolor{purple}{0} \\
\textcolor{purple}{0} & \textcolor{purple}{1} & \textcolor{purple}{0} & \textcolor{purple}{0} & \textcolor{purple}{0} & \textcolor{purple}{0} & \cdots & \textcolor{purple}{0} & \textcolor{purple}{0} \\
\textcolor{purple}{0} & \textcolor{purple}{0} & \textcolor{purple}{1} & \textcolor{purple}{0} & \textcolor{purple}{0} & \textcolor{purple}{0} & \cdots & \textcolor{purple}{0} & \textcolor{purple}{0} \\
\textcolor{purple}{0} & \textcolor{purple}{0} & \textcolor{purple}{0} & \textcolor{purple}{1} & \textcolor{purple}{0} & \textcolor{purple}{0} & \cdots & \textcolor{purple}{0} & \textcolor{purple}{0} \\
\vdots & \vdots & \vdots & \vdots & \vdots & \vdots & \cdots & \vdots & \vdots  \\
\textcolor{purple}{0} & \textcolor{purple}{0} & \textcolor{purple}{0} & \textcolor{purple}{0} & \textcolor{purple}{0} & \textcolor{purple}{0} & \cdots  & \textcolor{purple}{1} & \textcolor{purple}{0}\\
\textcolor{purple}{0} & \textcolor{purple}{0} & \textcolor{purple}{0} & \textcolor{purple}{0} & \textcolor{purple}{0} & \textcolor{purple}{0} & \cdots & \textcolor{purple}{0} & \textcolor{purple}{1} \\
\end{array}
\end{footnotesize} \right]}_{\bm{S}_t}
\underbrace{\left[
\begin{footnotesize}
\begin{array}{l}
\text{R}_{\text{P1*F},t} \\
\text{R}_{\text{P1*M},t} \\
\text{R}_{\text{P2*F},t} \\
\text{R}_{\text{P2*M},t} \\
\vdots \\
\text{R}_{\text{P47*F},t} \\
\text{R}_{\text{P47*M},t} \\
\end{array}
\end{footnotesize}
\right]}_{\bm{b}_t}
\]
\\

\noindent or $\bm{R}_t = \bm{S}_t\bm{b}_t$, where $\bm{R}_t$ is a vector containing all series at all levels of disaggregation, $\bm{b}_t$ is a vector of the most disaggregated series, and $\bm{S}_t$ shows how the two are connected. 

As evident in Figure~\ref{fig:2}, the group structure is not unique, since Japan can also be disaggregated by sex first. As a result of the non-uniqueness of the group structure, we consider the bottom-up and optimal combination methods of \cite{SH17b}, which are reviewed in Sections~\ref{sec:bu} and~\ref{sec:ols}, respectively. Their point and interval forecast accuracies are compared with the independent forecasts (without reconciliation) in Section~\ref{sec:6}. 

\subsection{Bottom-up method}\label{sec:bu}

The bottom-up method first generates independent forecasts for each series at the most disaggregated level and then aggregates these to produce all required forecasts. The bottom-up method performs well when the bottom-level series have a high signal-to-noise ratio. However, the bottom-up method may lead to inaccurate forecasts of the top-level series when the series contains missing or noisy data at the bottom level. 

\subsection{Optimal combination method}\label{sec:ols}

A disadvantage of the bottom-up method is that only the bottom-level series are considered. This disadvantage motivated \cite{HAA+11} to propose the optimal combination method, in which independent forecasts for all series are computed, and then the resultant forecasts are reconciled to satisfy the aggregation constraints via the summing matrix. The method is derived by expressing the independent forecasts as the response variable of the linear regression:
\begin{equation*}
\widehat{\bm{R}}_{n+h} = \bm{S}_{n+h}\bm{\beta}_{n+h}+\bm{\epsilon}_{n+h},
\end{equation*}
where $\widehat{\bm{R}}_{n+h}$ is a matrix of $h$-step-ahead independent forecasts for all series, stacked in the exact same order as for the original data; $\bm{\beta}_{n+h}=\text{E}[\bm{b}_{n+h}|\bm{R}_1,\dots,\bm{R}_n]$ is the unknown mean of the independent forecasts of the most disaggregated series; and $\bm{\epsilon}_{n+h}$ represents the forecast reconciliation errors.

To estimate the unknown regression coefficient, \cite{HAA+11} and \cite{HLW16} proposed a weighted least squares solution:
\begin{equation*}
\widehat{\bm{\beta}}_{n+h} = \left(\bm{S}_{n+h}^{\top}\bm{W}_{h}^{-1}\bm{S}_{n+h}\right)^{-1}\bm{S}_{n+h}^{\top}\bm{W}_{h}^{-1}\widehat{\bm{R}}_{n+h},
\end{equation*}
where $^{\top}$ denotes matrix transpose and $\bm{W}_{h}$ denotes a diagonal matrix. Assuming that $\bm{W}_h = k_h\bm{I}$ and $\bm{I}$ denotes identical matrix, then the revised forecasts are given by:
\begin{equation*}
\overline{\bm{R}}_{n+h} = \bm{S}_{n+h}\widehat{\bm{\beta}}_{n+h} = \bm{S}_{n+h}\left(\bm{S}_{n+h}^{\top}\bm{S}_{n+h}\right)^{-1}\bm{S}_{n+h}^{\top}\widehat{\bm{R}}_{n+h},
\end{equation*}
where $k_h$ is an arbitrary constant. These reconciled forecasts are aggregate consistent, involve a combination of all the independent forecasts, and are computationally fast. From the perspective of statistical properties, the reconciled forecasts are unbiased because $\text{E}(\widehat{\bm{\beta}}_{n+h})\rightarrow \bm{\beta}_{n+h}$ and $\text{E}(\overline{R}_{n+h})=\bm{S}_{n+h}\bm{\beta}_{n+h}$. Note that it is possible to consider the weighted least squares estimator of \cite{WAH19}, yet it may be computationally slower.

A crucial part of the improvement in forecast accuracy enjoyed by the reconciliation methods relies on the accurate forecast of the $S$ matrix. Recall that the $S$ matrix includes ratios of forecast exposure-to-risk. We make a reasonable cohort assumption: the observed ratios that form the $S$ matrix are forecast using the automatic autoregressive integrated moving average algorithm of \cite{HK08} for age $u_1=0$ \citep[c.f.,][]{SH17b}. For an age above 0, the exposure-to-risk of age $u_{i+1}$ in the year $t+1$ will be the same as the exposure-to-risk of age $u_i$ in the year $t$. Although we assume the population is closed (i.e., no external migration is allowed in Japan), this assumption is reasonable, and we only need to focus on the ratio of two exposure-to-risks. 

\section{Results: point forecasts}\label{sec:6}

\subsection{Functional time series model fitting}

\begin{figure}[!htbp]
\centering
\includegraphics[width = 7.3cm]{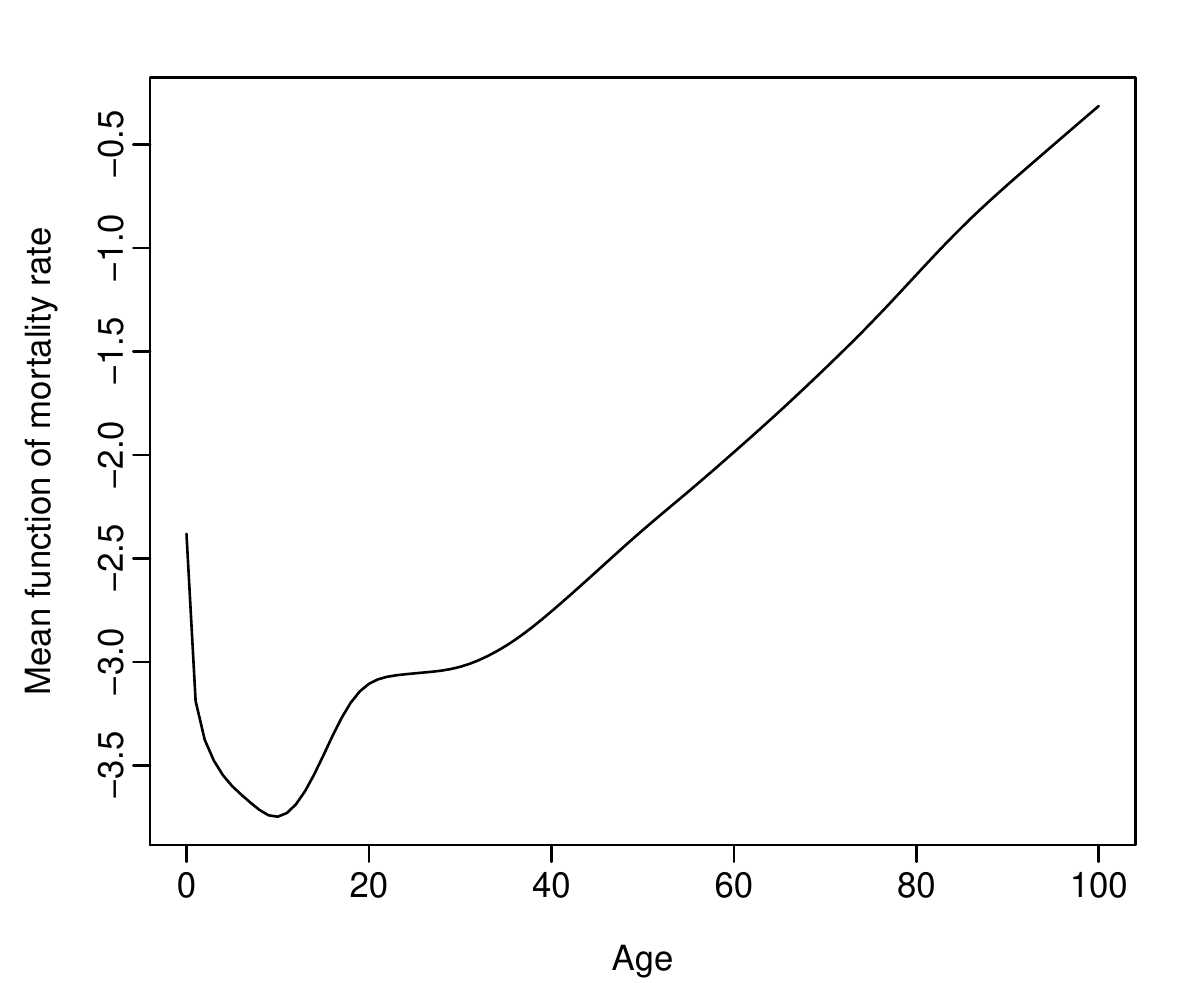}
\qquad
\includegraphics[width = 7.3cm]{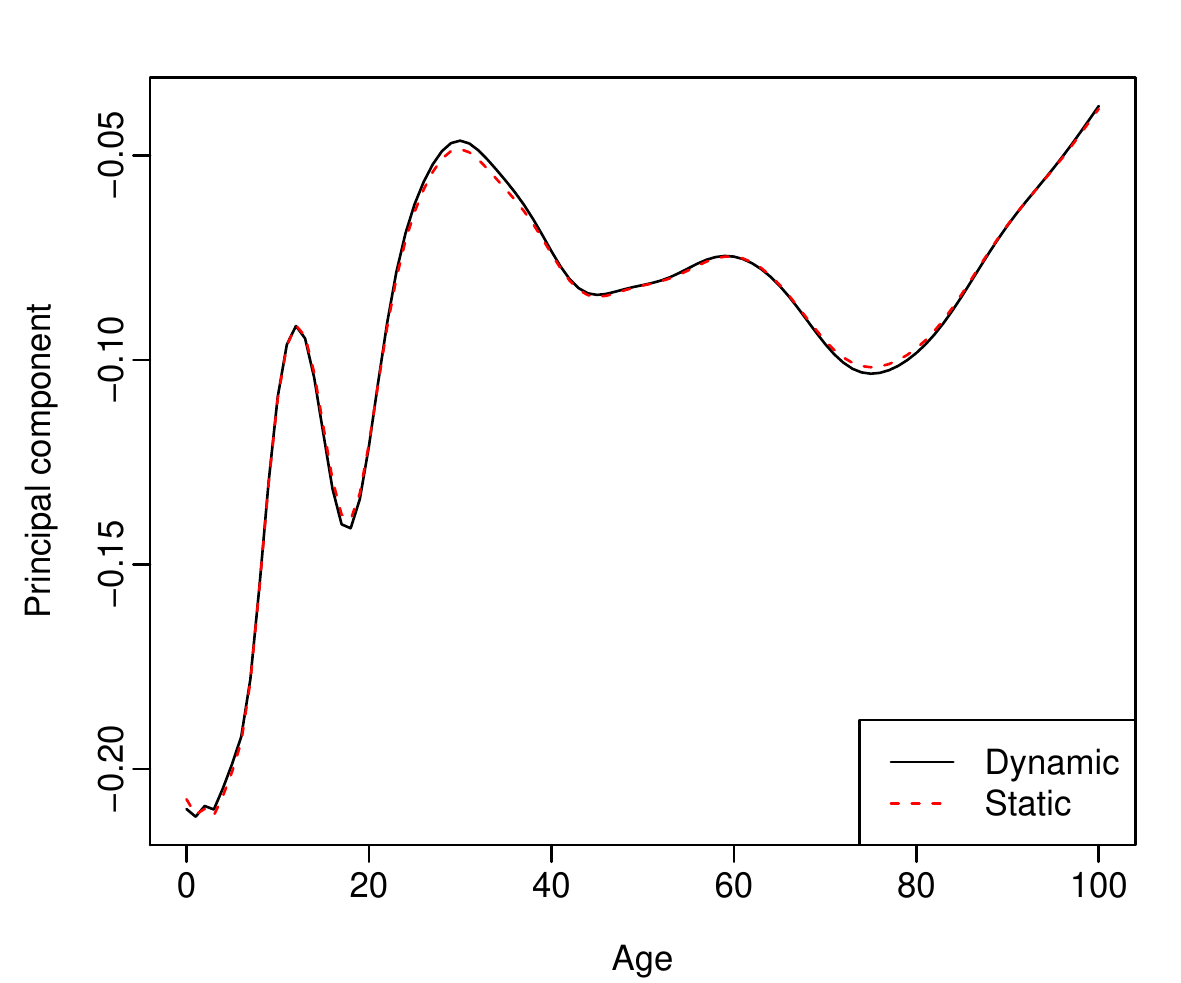}
\includegraphics[width = 7.3cm]{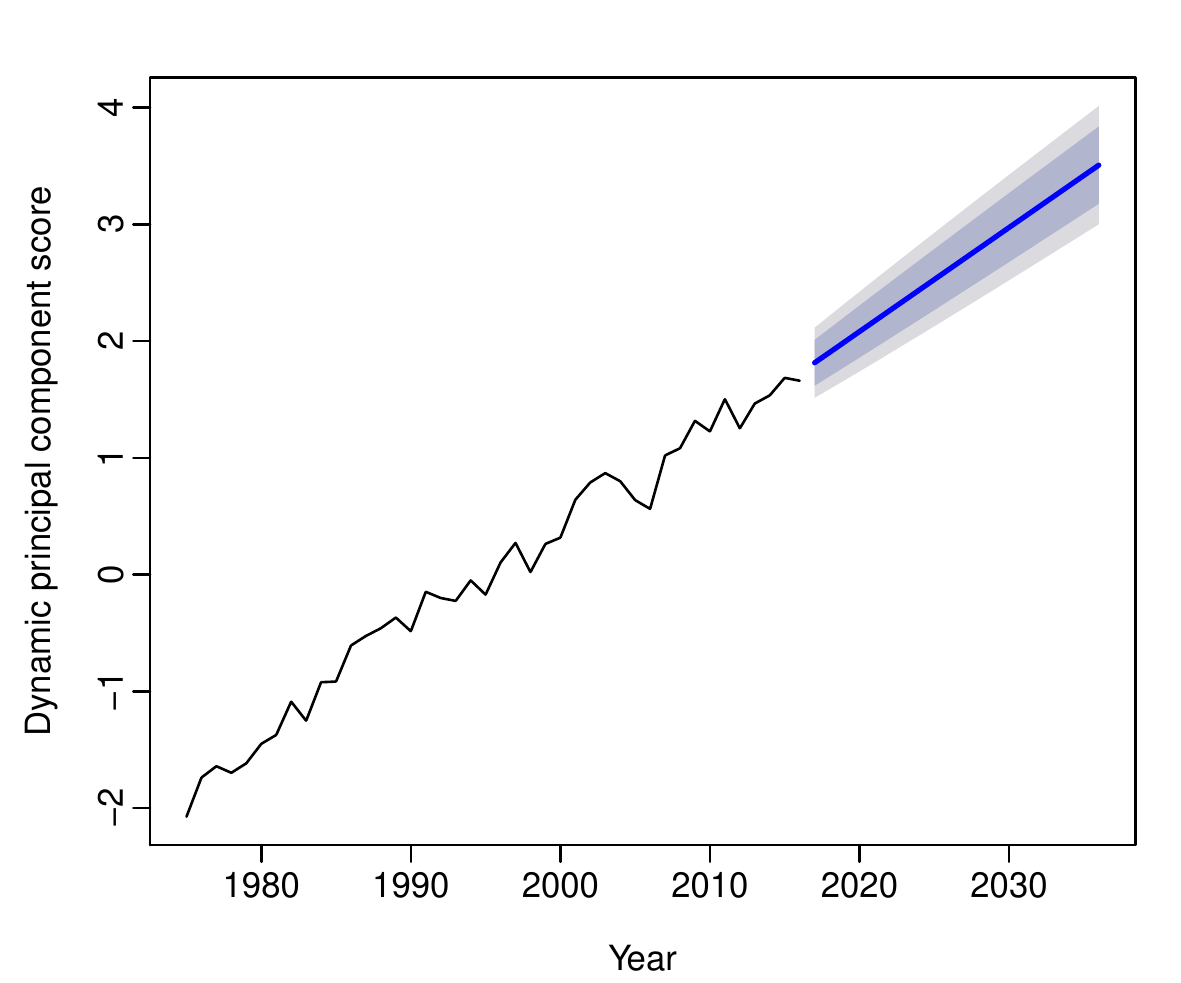}
\qquad
\includegraphics[width = 7.3cm]{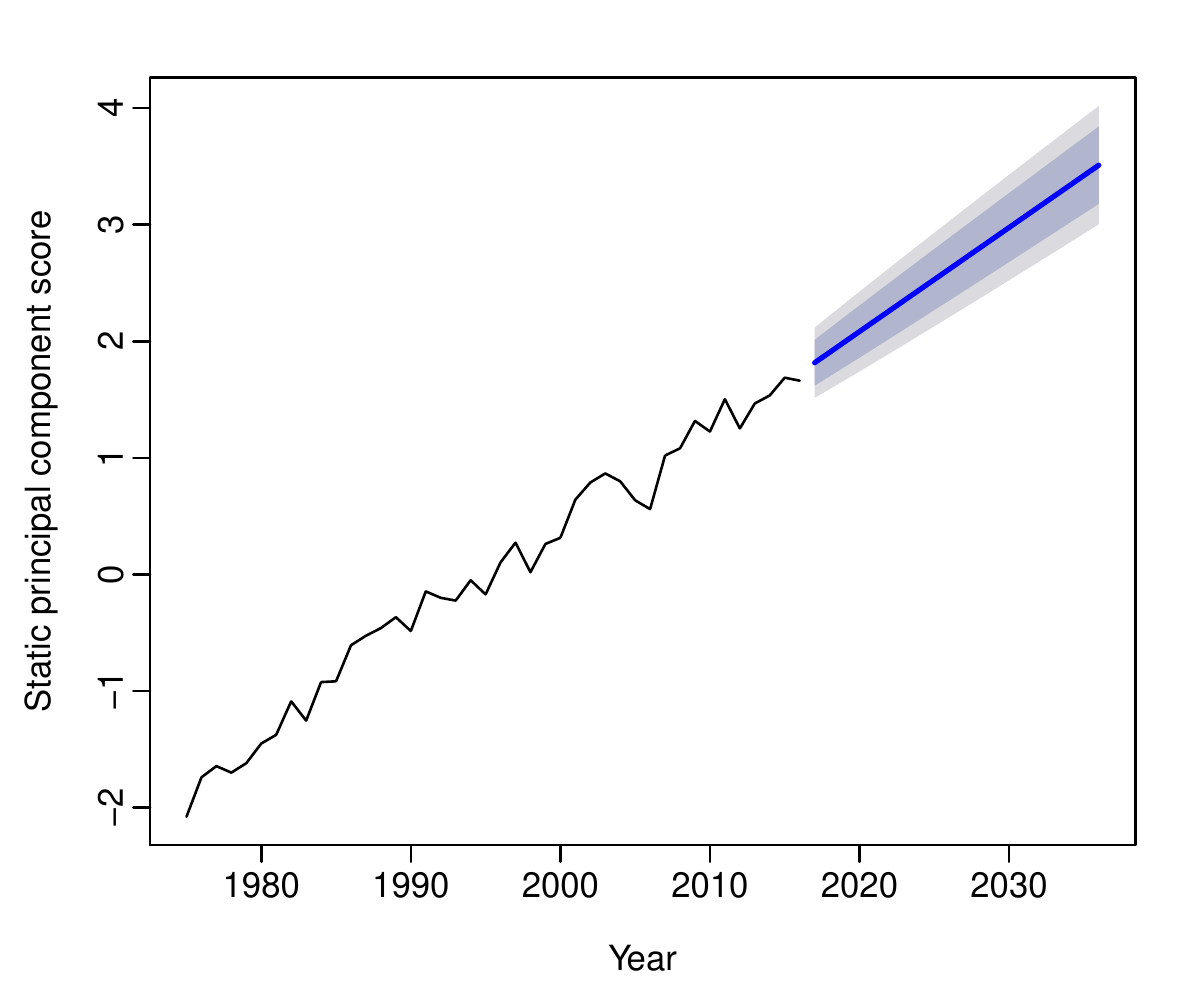}
\caption{Dynamic and static functional principal component decompositions for male log mortality rates in Hokkaido. The top panel shows the mean function and the first dynamic and static functional principal components, while the bottom panel shows the first dynamic and static functional principal component scores.}\label{fig:3}
\end{figure}

In the first row of Figure~\ref{fig:3}, we present the mean function for male log mortality rates in Hokkaido, and the first dynamic functional principal component, accounting for 95.4\% of the total variation, and its associated principal component scores. In the second row, we present the first static functional principal component, accounting for 90.4\% of the total variation, and its associated principal component scores. 

\begin{figure}[!htbp]
\centering
\subfloat[Dynamic functional principal component decomposition]
{\includegraphics[width=8.5cm]{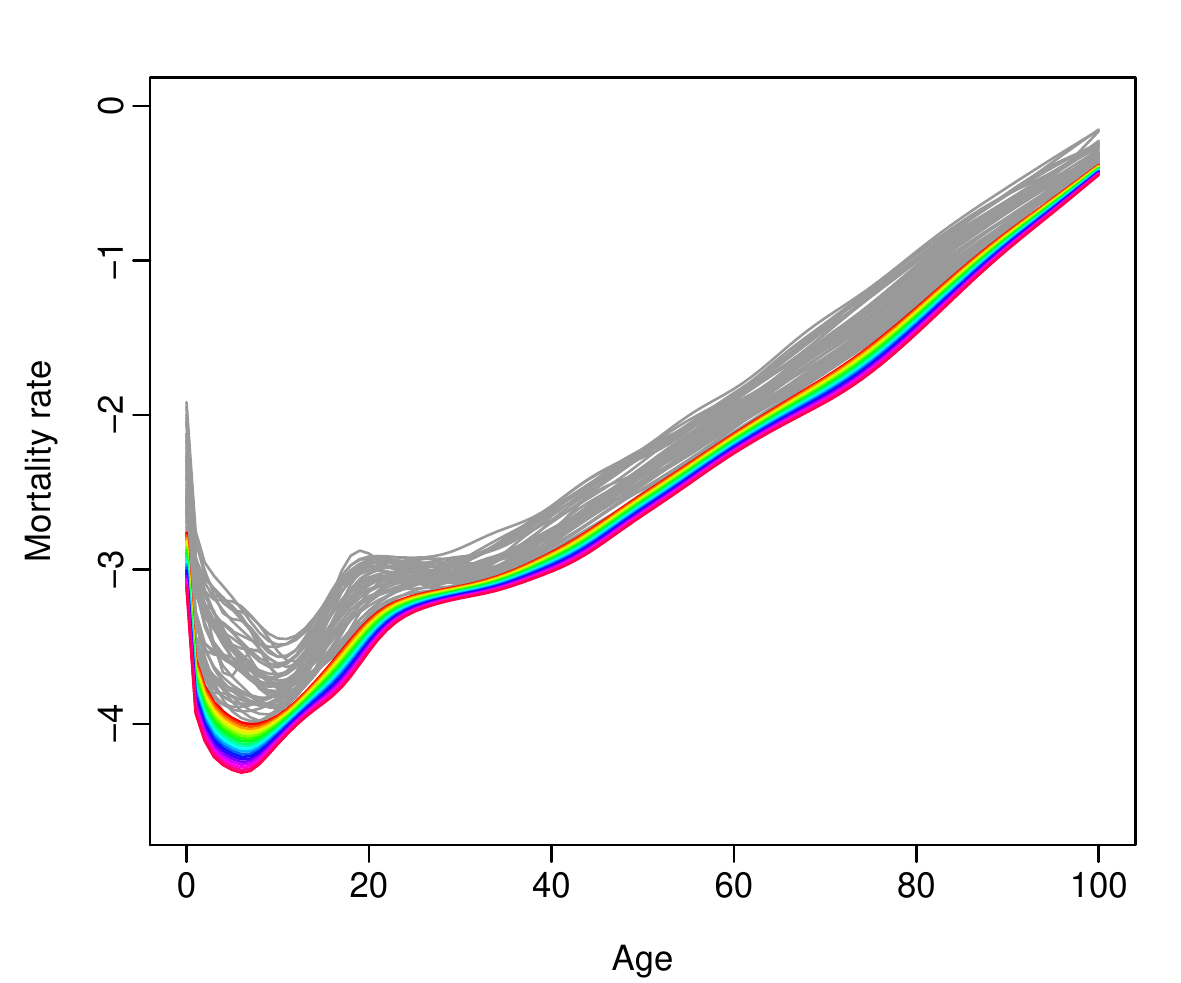}}
\qquad
\subfloat[Static functional principal component decomposition]
{\includegraphics[width=8.5cm]{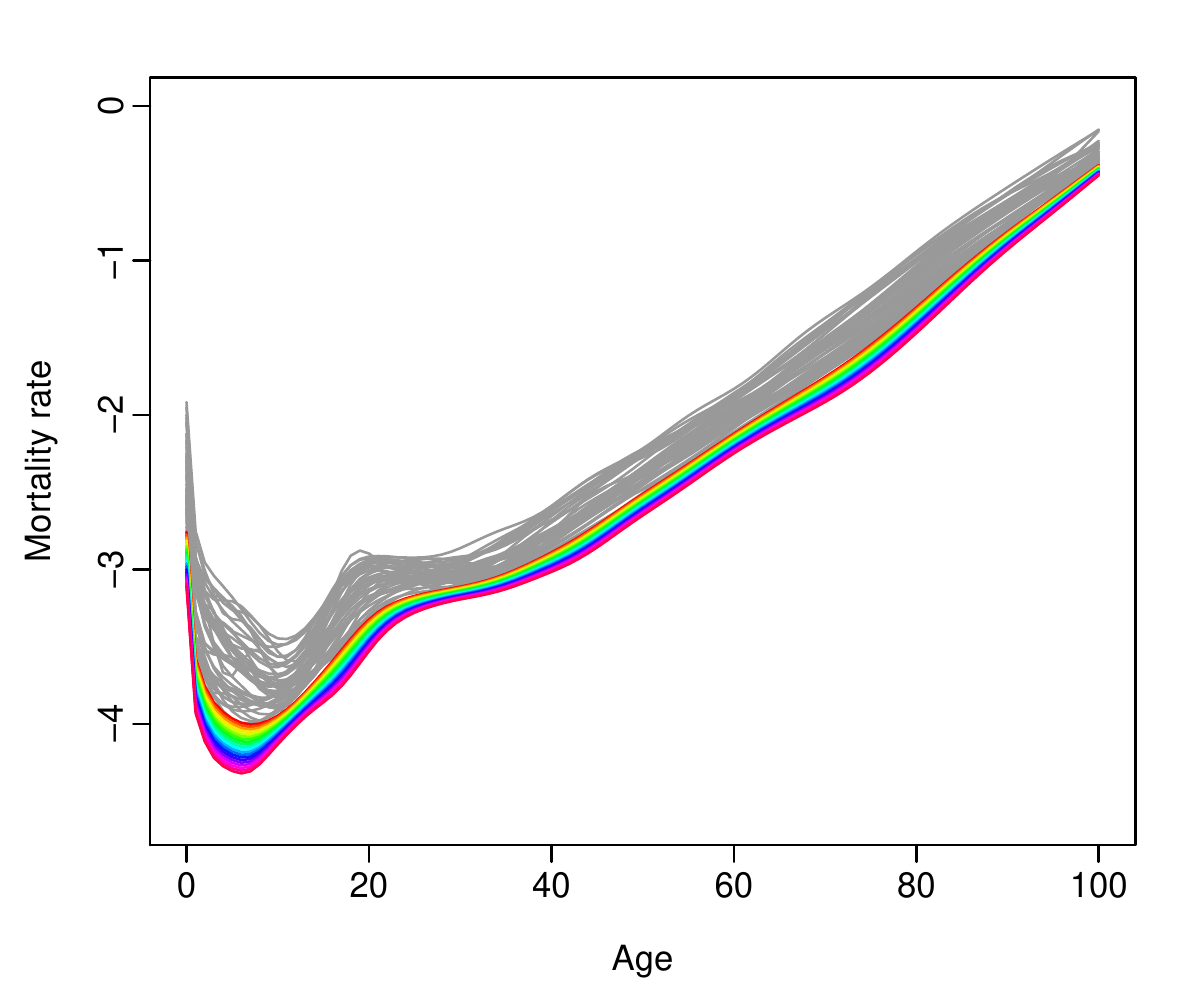}}
\caption{Point forecast of age-specific smoothed male mortality rates from 2017 to 2036 for the Japanese prefecture of Hokkaido. The historical smoothed functional time series is shown in gray and the forecasts are highlighted in rainbow colors.}\label{fig:4}
\end{figure}

Based on the historical mortality from 1975 to 2016, we produce the point forecasts of age-specific mortality rates from 2017 to 2036. As shown in Figure~\ref{fig:4}, the mortality rates are continuing to decline.

In Table~\ref{tab:2}, we tabulate the percentage of total variation explained by the first dynamic or static functional principal component for the national and subnational smoothed log mortality rates. The first dynamic functional principal component always explains more variability than does the first static functional principal component. 

\begin{singlespace}
\begin{center}
\begin{small}
\tabcolsep 0.09in
\begin{longtable}{@{}lccccllcccc@{}}
\caption{Percentage of total variation explained by the first dynamic and the first static functional principal component decompositions for fitting smoothed log mortality. The static functional principal component analysis is abbreviated as FPCA, while the dynamic functional principal component analysis is abbreviated as DFPCA.}\label{tab:2}\\
\toprule
 & \multicolumn{2}{c}{Female} & \multicolumn{2}{c}{Male} & & & \multicolumn{2}{c}{Female} & \multicolumn{2}{c}{Male} \\
 Prefecture 		& FPCA & DFPCA & FPCA  & DFPCA & & Prefecture & FPCA & DFPCA  & FPCA  & DFPCA  \\
\midrule
\endfirsthead
\toprule
 & \multicolumn{2}{c}{Female} & \multicolumn{2}{c}{Male} & & & \multicolumn{2}{c}{Female} & \multicolumn{2}{c}{Male} \\
 Prefecture 		& FPCA & DFPCA & FPCA  & DFPCA & & Prefecture & FPCA & DFPCA  & FPCA  & DFPCA  \\
\midrule
\endhead
\hline \multicolumn{11}{r}{{Continued on next page}} \\ 
\endfoot
\endlastfoot
  Japan & 0.969 & \textBF{0.981} & 0.965 & \textBF{0.974} & & Mie &  0.849 & \textBF{0.949} & 0.778 & \textBF{0.893} \\ 
  Hokkaido & 0.895 & \textBF{0.954} & 0.904 & \textBF{0.954} & & Shiga & 0.876 & \textBF{0.962} & 0.771 & \textBF{0.898} \\ 
  Aomori & 0.840 & \textBF{0.936} & 0.792 & \textBF{0.907} & & Kyoto & 0.897 & \textBF{0.956} & 0.823 & \textBF{0.932} \\ 
  Iwate & 0.714 & \textBF{0.858} & 0.735 & \textBF{0.837} & & Osaka &  0.933 & \textBF{0.967} & 0.913 & \textBF{0.954} \\ 
  Miyagi & 0.745 & \textBF{0.829} & 0.760 & \textBF{0.858} & & Hyogo &  0.873 & \textBF{0.956} & 0.893 & \textBF{0.958} \\ 
  Akita & 0.788 & \textBF{0.905} & 0.763 & \textBF{0.905} & & Nara &  0.870 & \textBF{0.952} & 0.795 & \textBF{0.906} \\ 
  Yamagata & 0.837 & \textBF{0.931} & 0.729 & \textBF{0.877} & & Wakayama & 0.735 & \textBF{0.878} & 0.690 & \textBF{0.851} \\ 
  Fukushima & 0.813 & \textBF{0.935} & 0.781 & \textBF{0.893} & & Tottori &  0.770 & \textBF{0.906} & 0.699 & \textBF{0.859} \\ 
  Ibaraki & 0.863 & \textBF{0.953} & 0.841 & \textBF{0.927} & & Shimane & 0.805 & \textBF{0.927} & 0.714 & \textBF{0.873} \\  
  Tochigi & 0.874 & \textBF{0.952} & 0.786 & \textBF{0.905} & & Okayama &  0.893 & \textBF{0.971} & 0.824 & \textBF{0.932} \\ 
  Gunma & 0.872 & \textBF{0.950} & 0.817 & \textBF{0.932} & & Hiroshima & 0.891 & \textBF{0.959} & 0.846 & \textBF{0.920} \\ 
  Saitama & 0.915 & \textBF{0.963} & 0.887 & \textBF{0.945} & & Yamaguchi & 0.815 & \textBF{0.938} & 0.755 & \textBF{0.889} \\ 
  Chiba & 0.924 & \textBF{0.968} & 0.863 & \textBF{0.942} & & Tokushima &  0.819 & \textBF{0.931} & 0.791 & \textBF{0.917} \\ 
  Tokyo & 0.936 & \textBF{0.969} & 0.915 & \textBF{0.952} & & Kagawa & 0.772 & \textBF{0.928} & 0.718 & \textBF{0.904} \\ 
  Kanagawa & 0.915 & \textBF{0.962} & 0.885 & \textBF{0.945} & & Ehime & 0.829 & \textBF{0.929} & 0.742 & \textBF{0.876} \\ 
  Niigata & 0.883 & \textBF{0.955} & 0.865 & \textBF{0.954} & & Kochi & 0.754 & \textBF{0.881} & 0.771 & \textBF{0.914} \\ 
  Toyama & 0.841 & \textBF{0.951} & 0.737 & \textBF{0.884} & & Fukuoka & 0.902 & \textBF{0.960} & 0.901 & \textBF{0.956} \\ 
  Ishikawa & 0.820 & \textBF{0.938} & 0.764 & \textBF{0.901} & & Saga & 0.851 & \textBF{0.950} & 0.750 & \textBF{0.883} \\ 
  Fukui & 0.779 & \textBF{0.920} & 0.751 & \textBF{0.893} & & Nagasaki & 0.848 & \textBF{0.950} & 0.827 & \textBF{0.935} \\ 
  Yamanashi & 0.806 & \textBF{0.920} & 0.756 & \textBF{0.901} & & Kumamoto & 0.871 & \textBF{0.946} & 0.846 & \textBF{0.950} \\ 
  Nagano & 0.865 & \textBF{0.947} & 0.800 & \textBF{0.932} & & Oita & 0.848 & \textBF{0.942} & 0.808 & \textBF{0.921} \\ 
  Gifu & 0.869 & \textBF{0.962} & 0.795 & \textBF{0.911} & & Miyazaki & 0.819 & \textBF{0.926} & 0.772 & \textBF{0.926} \\ 
  Shizuoka & 0.891 & \textBF{0.961} & 0.858 & \textBF{0.933} & & Kagoshima & 0.875 & \textBF{0.948} & 0.860 & \textBF{0.946} \\ 
  Aichi & 0.938 & \textBF{0.976} & 0.897 & \textBF{0.949} & & Okinawa & 0.834 & \textBF{0.925} & 0.762 & \textBF{0.890} \\ 
  \bottomrule
\end{longtable}
\end{small}
\end{center}
\end{singlespace}

\newpage
\subsection{Point forecast evaluation}

An expanding window approach of a time series model is commonly used to assess model and parameter stabilities over time, and prediction accuracy. The expanding window approach assesses the constancy of a model's parameter by computing parameter estimates and their resultant forecasts over an expanding window of a fixed size through the sample \citep[for details, see][pp. 313-314]{ZW06}. Using the first 12 observations from 1975 to 1986 in the Japanese age-specific mortality rates, we produce one- to 30-step-ahead point forecasts. Through a rolling window approach, we re-estimate the parameters in the time series forecasting models using the first 13 observations from 1975 to 1987. Forecasts from the estimated models are then produced for one- to 30-step-ahead. We iterate this process by increasing the sample size by one year until reaching the end of the data period in 2016. This process produces 30 one-step-ahead forecasts, 29 two-step-ahead forecasts, $\dots$, and one 30-step-ahead forecast. We compare these forecasts with the holdout samples to determine the out-of-sample point forecast accuracy. 

To evaluate the point forecast accuracy, we consider mean absolute forecast error (MAFE) and root mean squared forecast error (RMSFE). These criteria measure how close forecasts are to the actual values of the raw mortality rate being forecast, regardless of the direction of forecast errors. For each series, $k$, these error measures can be expressed as:
\begin{align*}
\text{MAFE}_k(h) &= \frac{1}{101\times (31-h)}\sum^{30}_{\xi=h}\sum^{101}_{i=1}\left|\X_{m+\xi}^k(u_i) - \widehat{\X}_{m+\xi}^k(u_i)\right|, \\
\text{RMSFE}_k(h) &= \sqrt{\frac{1}{101\times (31-h)}\sum^{30}_{\xi=h}\sum^{101}_{i=1}\left[\X_{m+\xi}^k(u_i) - \widehat{\X}_{m+\xi}^k(u_i)\right]^2}
\end{align*}
where $m$ denotes the last year of the fitting period, $\X_{m+\xi}^k(u_i)$ denotes the actual holdout sample for the $i$\textsuperscript{th} age and $\xi$\textsuperscript{th} curve of the forecasting period in the $k$\textsuperscript{th} series, and $\widehat{\X}_{m+\xi}^k(u_i)$ denotes the point forecasts for the holdout sample.

By averaging MAFE$_k(h)$ and RMSFE$_k(h)$ across the number of series within each level of disaggregation, we obtain an overall assessment of the point forecast accuracy for each level within the collection of series, denoted by MAFE$(h)$ and RMSFE$(h)$. These error measures are defined as:
\begin{align*}
\text{MAFE}(h) &= \frac{1}{m_k}\sum^{m_k}_{k=1}\text{MAFE}_k(h), \\
\text{RMSFE}(h) &= \frac{1}{m_k}\sum^{m_k}_{k=1}\text{RMSFE}_k(h),
\end{align*}
where $m_k$ denotes the number of series at the $k$\textsuperscript{th} level of disaggregation, for $k=1,\dots,K$. In the Japanese group structure, $K=6$.

For 30 different forecast horizons, we consider two summary statistics to evaluate overall point forecast accuracy. The chosen summary statistics are the mean and median values because of their suitability for handling squared and absolute errors \citep{Gneiting11}. These error measures are given by
\begin{align*}
\text{Mean (RMSFE)} &= \frac{1}{30}\sum^{30}_{h=1}\text{RMSFE}(h), \\
\text{Median (MAFE)} &= \frac{1}{2}\left(\text{MAFE}[15] + \text{MAFE}[16]\right),
\end{align*}
where $[\cdot]$ denotes an ordered statistic.

\subsection{Comparison of point forecast accuracy}

\begin{figure}[!htbp]
\centering
\includegraphics[width=\textwidth]{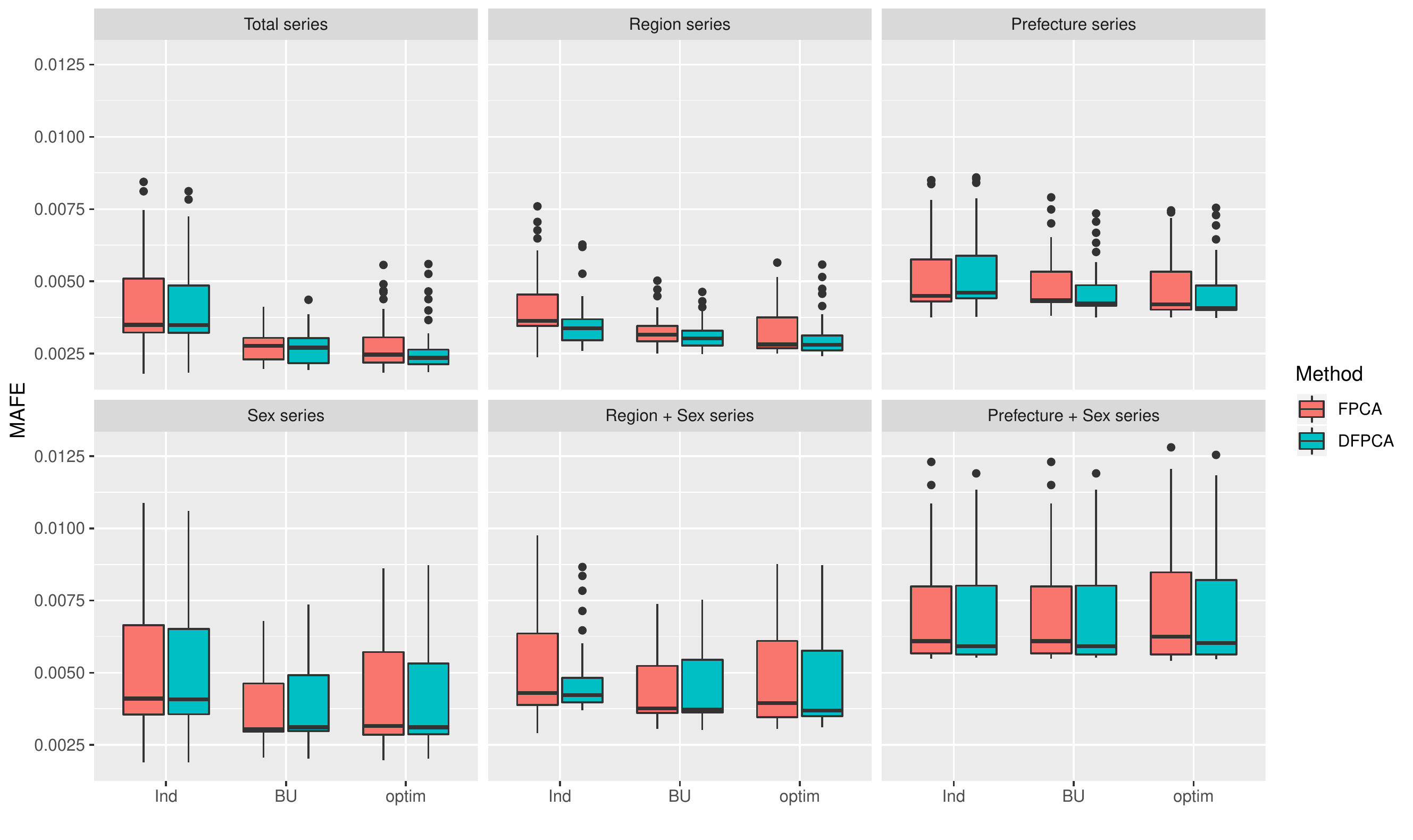}
\caption{MAFE comparison of the independent and reconciled forecasts obtained from the functional time series method with static and dynamic functional principal component decompositions.}\label{fig:MAFE}
\end{figure}

\begin{figure}[!htbp]
\centering
\includegraphics[width=\textwidth]{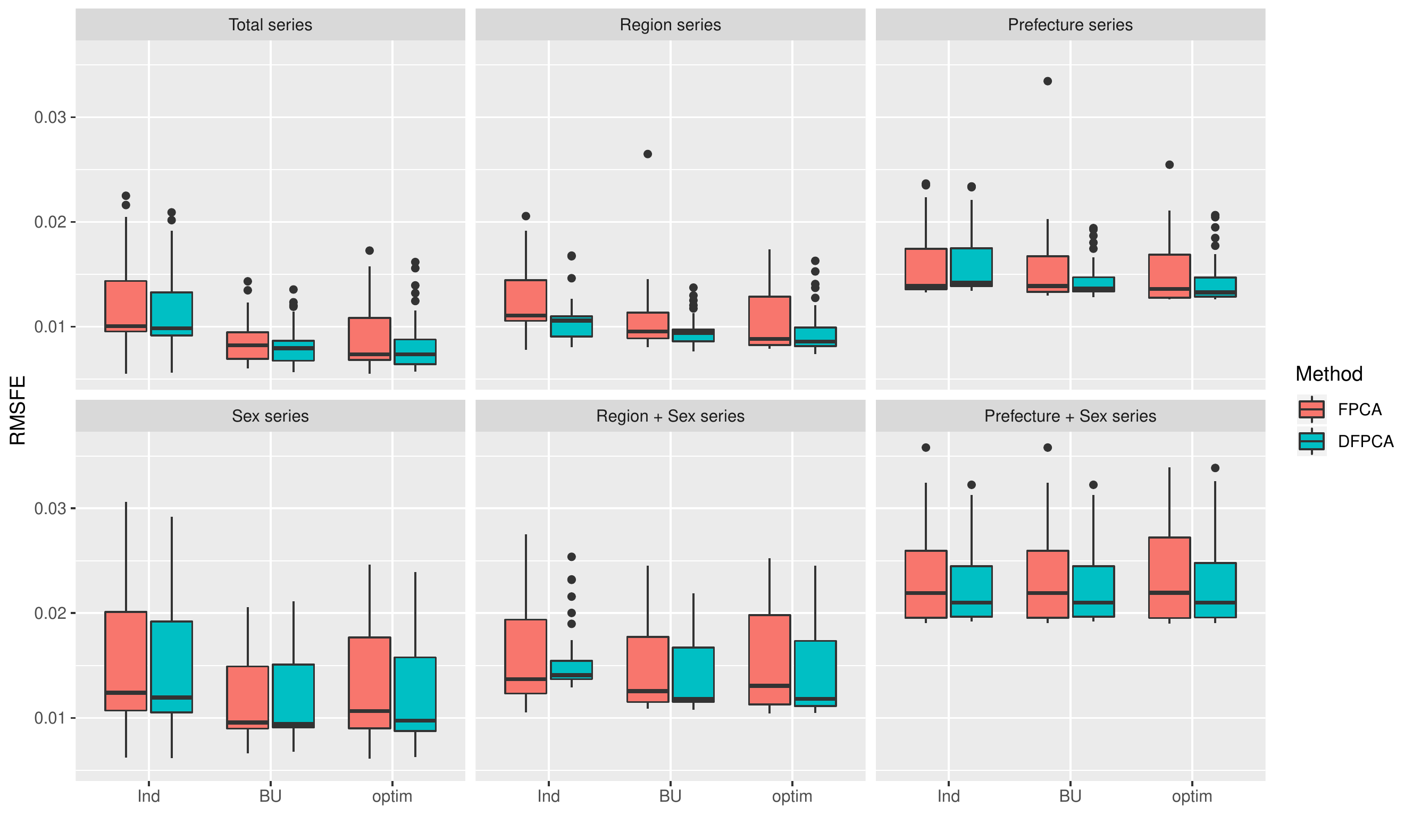}
\caption{RMSFE comparison of the independent and reconciled forecasts obtained from the functional time series method with static and dynamic functional principal component decompositions.}\label{fig:RMSFE}
\end{figure}

Through averaging over all the series at each level of the group structure, Figures~\ref{fig:MAFE} and~\ref{fig:RMSFE} present the one-step-ahead to 30-step-ahead MAFE and RMSFE of the independent forecasts obtained from the functional time series method with the static and dynamic functional principal component decompositions. Given that we evaluate and assess the point forecast accuracy for 30 different forecast horizons, we consider two summary statistics. The median is ideal for absolute error, while the mean is ideal for squared error \citep{Gneiting11}. From the MAFE and RMSFE measures, more accurate point forecasts can be obtained by the functional time series method with the dynamic functional principal component decomposition, at each level of the group structure. The forecast accuracy of the functional time series method with dynamic functional principal component decomposition is superior because it better captures the temporal dependence in each series. 

\section{Results: interval forecasts}\label{sec:int}

\subsection{Interval forecast evaluation}

To evaluate pointwise interval forecast accuracy, we use the interval score of \cite{GR07} and \cite{GK14}. For each year in the forecasting period, the $h$-step-ahead pointwise prediction intervals are computed at the $100(1-\alpha)\%$ nominal coverage probability. We consider the symmetric prediction intervals, with lower and upper bounds that are predictive quantiles at $\alpha/2$ and $1-\alpha/2$, denoted by $\widehat{\X}^{k, \text{lb}}_{m+\xi}(u_i)$ and $\widehat{\X}^{k, \text{ub}}_{m+\xi}(u_i)$. As defined by \cite{GR07}, a scoring rule for the interval forecasts at time point $\X^k_{m+\xi}(u_i)$ for series $k$ is:
\begin{align*}
S_{\alpha}\left[\widehat{\X}_{m+\xi}^{k, \text{lb}}(u_i), \widehat{\X}_{m+\xi}^{k, \text{ub}}(u_i); \X^k_{m+\xi}(u_i)\right]=&\left[\widehat{\X}^{k,\text{ub}}_{m+\xi}(u_i) - \widehat{\X}^{k,\text{lb}}_{m+\xi}(u_i)\right]+\\
&\frac{2}{\alpha}\left[\widehat{\X}_{m+\xi}^{k,\text{lb}}(u_i) - \X^k_{m+\xi}(u_i)\right] \mathds{1}\left\{\X^k_{m+\xi}(u_i)<\widehat{\X}_{m+\xi}^{k, \text{lb}}(u_i)\right\}+\\
&\frac{2}{\alpha}\left[\X^k_{m+\xi}(u_i) - \widehat{\X}_{m+\xi}^{k, \text{ub}}(u_i)\right]\mathds{1}\left\{\X^k_{m+\xi}(u_i)>\widehat{\X}_{m+\xi}^{k, \text{ub}}(u_i)\right\}
\end{align*}
where $\mathds{1}\{\cdot\}$ represents the binary indicator function, and $\alpha$ denotes the level of significance, customarily $\alpha = 0.2$. 

Given that more samples are needed to compute the discrepancy between the in-sample holdout curves and their corresponding forecast curves in~\eqref{eq:4}, we consider 15 forecast horizons instead of 30 forecast horizons, as was the case for evaluating point forecast accuracy. Through averaging across ages and years in the forecasting period, we obtain the mean interval score, expressed as:
\begin{equation*}
\overline{S}^k_{\alpha}(h) = \frac{1}{101\times (16-h)}\sum^{15}_{\xi=h}\sum^{101}_{i=1}S_{\alpha}\left[\widehat{\X}_{m+\xi}^{k, \text{lb}}(u_i), \widehat{\X}_{m+\xi}^{k, \text{ub}}(u_i); \X^k_{m+\xi}(u_i)\right].
\end{equation*}
The mean interval score rewards a narrow prediction interval if, and only if, the actual observation lies within the prediction interval. 

By averaging $\overline{S}^k_{\alpha}(h)$ across the number of series within each level of disaggregation, we obtain an overall assessment of the interval forecast accuracy for each level within the collection of series, denoted by $\overline{S}_{\alpha}(h)$. The error measure is defined as:
\begin{equation*}
\overline{S}_{\alpha}(h) = \frac{1}{m_k}\sum^{m_k}_{k=1}\overline{S}^k_{\alpha}(h),
\end{equation*}
where $m_k$ denotes the number of series at the $k$\textsuperscript{th} level of disaggregation, for $k=1,\dots,K$. 

For 15 different forecast horizons, we consider two summary statistics to evaluate overall interval forecast accuracy. The mean and median of $\overline{S}_{\alpha}(h)$ are given by:
\begin{align*}
\overline{S}_{\alpha} &= \frac{1}{15}\sum^{15}_{h=1}\overline{S}_{\alpha}(h), \\
\text{Median}(\overline{S}_{\alpha}) &= \overline{S}_{\alpha}[8].
\end{align*}

\subsection{Comparison of interval forecast accuracy}

\begin{figure}[!htbp]
\centering
\includegraphics[width=\textwidth]{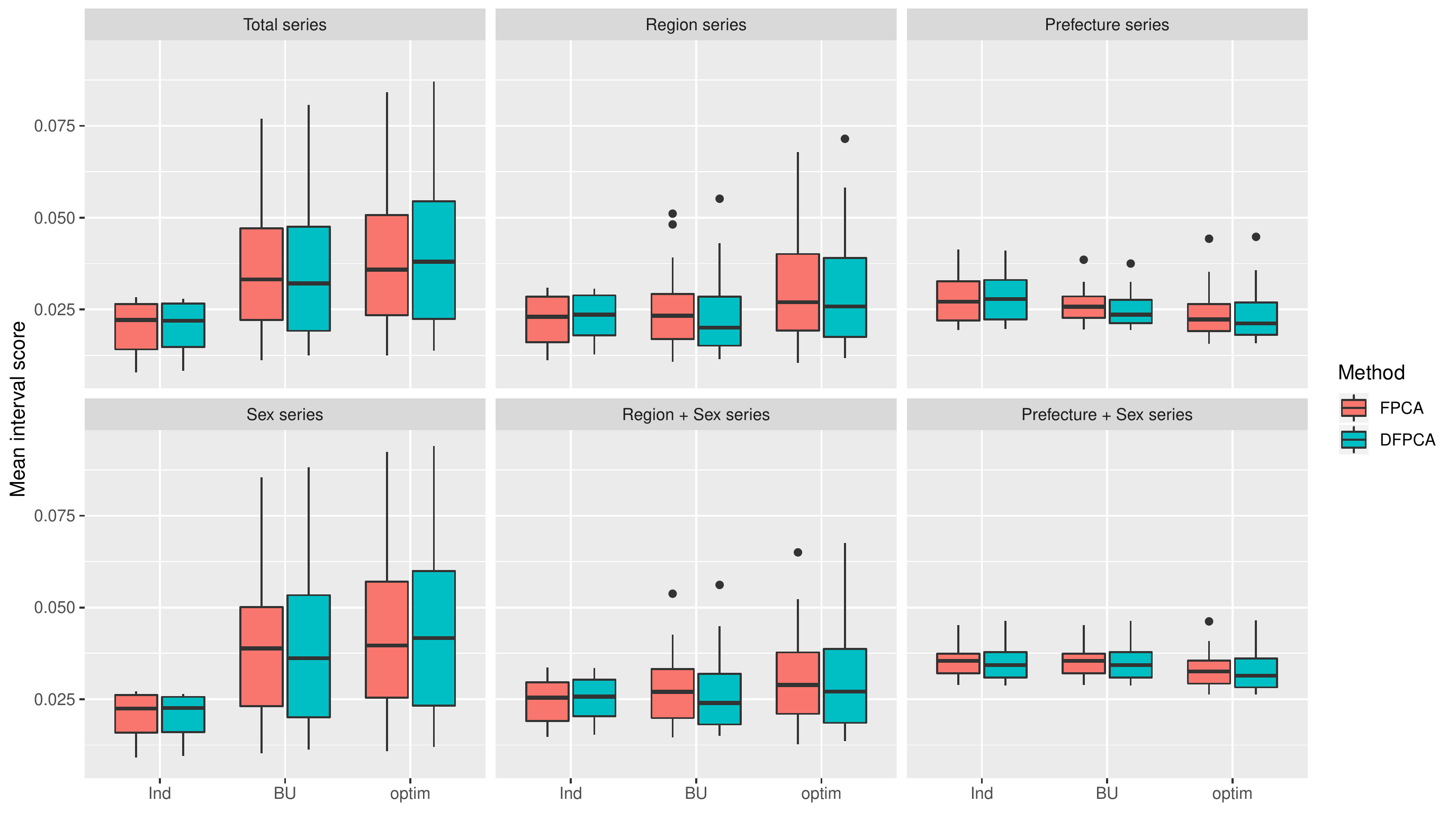}
\caption{Mean interval score comparison of the independent and reconciled forecasts obtained from the functional time series method with static and dynamic functional principal component decompositions.}\label{fig:score}
\end{figure}

Through averaging over all the series at each level of the group structure, Figure~\ref{fig:score} presents the mean interval scores and their summary statistics for the independent forecasts obtained from the functional time series method with the static and dynamic functional principal component decompositions. As also found in \cite{SH17b}, the independent forecasting method produces the smallest mean interval scores at the top level, when the data are often of higher quality. Given that the grouped forecasting methods account for correlation between series, they can potentially improve interval forecast accuracy, especially at the bottom level. Improvement in interval forecast accuracy at the subnational level is essential for decision making at the regional level. When grouped forecasting methods are considered, the forecast accuracy of the functional time series method with DFPCA is generally superior--it better captures the temporal dependence in each series, especially at the short-term forecast horizon. Between the two grouped forecasting methods, the optimal combination method produces the smallest mean interval scores at the prefecture level, yet not at the regional and national levels.

With the independent forecasting method, we observe that the functional time series method with dynamic functional principal component decomposition does not always outperform that with static functional principal component decomposition. From~\eqref{eq:4} and~\eqref{eq:5}, the difference in the in-sample errors between the static and dynamic functional principal component decompositions is absorbed when computing their corresponding interval forecasts.

\section{Conclusion}\label{sec:7}

Using national and subnational Japanese mortality data, we have evaluated and compared the point and interval forecast accuracies between the functional time series method with static and dynamic functional principal component analyses. Based on the forecast errors, the functional time series method with dynamic functional principal component decomposition generally outperforms that with static functional principal component decomposition. The superiority of the former is driven by the addition of auto-covariance in the estimated long-run covariance function. 

We implemented a functional time series method to produce independent forecasts. We then considered the issue of forecast reconciliation by applying two grouped functional time series forecasting methods--namely, the bottom-up and optimal combination methods. The bottom-up method models and forecasts data series at the most disaggregated level, and then aggregates the forecasts using the summing matrix constructed by forecast exposure-to-risk. The optimal combination method combines the independent forecasts obtained from independent functional time series forecasting methods using linear regression. The optimal combination method generates a set of revised forecasts that are as close as possible to the independent forecasts, yet also aggregate consistently with the known grouping structure. Under some mild technical assumptions, the regression coefficient may be estimated by ordinary least squares.

Illustrated by the Japanese age-specific mortality data, we examined forecast accuracy. In particular, we compared one-step-ahead to 30-step-ahead point forecast accuracy, and one-step-ahead to 15-step-ahead interval forecast accuracy between the independent forecasting method and two grouped forecasting methods. In terms of point forecast accuracy, the grouped functional time series forecasting methods produce more accurate point forecasts than do those obtained by the independent forecasting method, averaged across all levels of the group structure. In terms of interval forecast accuracy, the grouped forecasting methods produce more accurate interval forecasts than does the independent forecasting method at the prefecture level, yet not at the regional and national levels. Between the two grouped forecasting methods, they perform on par for producing point forecasts. However, the optimal combination method is recommended for producing interval forecasts at least for the data we considered. The grouped forecasting methods produce forecasts that obey the original group structure, resulting in the forecast mortality rates at the subnational level adding up to the forecast mortality rates at the national level. 

\newpage
\bibliographystyle{agsm}
\bibliography{GFTS_dynamic}

\end{document}